\documentclass[letterpaper, 10 pt, conference]{ieeeconf}  

\IEEEoverridecommandlockouts                              

\usepackage{amsmath}
\usepackage{amssymb}  
\usepackage{graphicx}
\title{\LARGE \bf
Adaptive Control of Dubins Vehicle in the Presence of Loss of Effectiveness
}

\author{Daniel Maldonado Naranjo and Anuradha Annaswamy
\thanks{This work was supported by the Boeing Strategic University Initiative. The first author would like to acknowledge the support of the National GEM consortium and the Thomas W. Folger (1948) Fellowship.
Both authors (\{d8maldon,aanna\}@mit.edu) are with the Department of Mechanical Engineering, Massachusetts
Institute of Technology, Cambridge, MA 02139 USA. Several useful discussions with Eugene Lavretsky and Heather Hussain are gratefully acknowledged.}}

\begin{document}

\maketitle
\thispagestyle{empty}
\pagestyle{empty}

\begin{abstract}
The control of a Dubins Vehicle when subjected to a loss of control effectiveness in the turning rate is considered. A complex state-space representation is used to model the vehicle dynamics. An adaptive control design is proposed, with the underlying stability analysis guaranteeing closed-loop boundedness and tracking of a desired path. It is shown that a path constructed by waypoints and a minimum turn radius can be specified using a reference model which can be followed by the closed loop system. The control design utilizes the complex state-space representation as well as a PID controller for the nominal closed-loop. How the design can be modified to ensure path following even in the presence input constraints is also discussed. Simulation studies are carried out to complement the theoretical derivations.
\end{abstract}

\section{Introduction}\label{sec:introduction}
The Dubins vehicle (DV) represents a canonical model of a vehicle that can be used for designing control methods to solve path-following and waypoint guidance problems in several applications. Recent areas of research aim to explore and expand upon Dubins' original work \cite{dubins1957curves}, applying its principles to a wide range of applications in aerospace, robotics, and marine vehicles \cite{frazzoli2002real, DubinsTSP2023, karapetyan2018multi, hernandez2019online}. 

As the DV model is nonlinear, standard approaches \cite{jha2018robust, anisi2003optimal, paden2016survey}  consider a linearized model followed by linear  control techniques such as Proportional-Integral-Derivative (PID) or Linear–Quadratic Regulator (LQR). An interesting approach is presented in \cite{2024JGCD...47.1256L} where the underlying bilinearity of the vehicle model is preserved by using a complex state-space representation which allows decoupling of the speed and turning rate into two separate control problems.
Path following can be accomplished using virtual target point (VTP) \cite{rysdyk2006unmanned, medagoda2010synthetic}, or vector field methods \cite{5504176,wang2022fixed} using waypoints. In this paper, we carry out path following for a DV using VTP methods when uncertainties are present in the form of loss of control effectiveness in the turning rate.

A sudden or gradual loss of effectiveness (LOE) in control actuators can occur at any time, highlighting the importance of developing control designs capable of maintaining operational integrity in the presence of parametric uncertainties. Over the past decades, the field of adaptive control (AC) has evolved from early developments such as the MIT rule to the formulation of model reference adaptive control (MRAC) and self-tuning regulators for deterministic and stochastic systems in both continuous and discrete time  \cite{narendra2012stable, annaswamy2021historical}. Recent advances focus on robustness, unmodeled dynamics, and handling input and state constraints all of which further enhance the practical applicability of adaptive control in modern engineering applications \cite{karason1993adaptive, lavretsky2024robust, ioannou1996robust}. Applying adaptive control techniques to quadrotor UAVs has demonstrated satisfactory performance even in the presence of loss of thrust due to actuator anomalies or other modeling errors \cite{6220873, whitehead2010model, nicol2011robust}. However, the underlying vehicle model is assumed to be a point mass, neglecting the constraints imposed by the physical aspects of the system. Recognizing this, it becomes essential to incorporate constraints such as a minimum turning radius to capture realistic maneuvering dynamics. In the DV, input constraints on speed and turning rate are introduced to accurately capture these limitations. These constraints limit the control commands to feasible ranges, ensuring that the system operates within safe and realistic bounds.

This work leverages a complex state-space representation of the
DV proposed in  \cite{2024JGCD...47.1256L} as opposed to the commonly used trigonometric representation.
Such a representation is beneficial as the
number of states are reduced by half, no approximation or
linearization is required in the control design, speed and
turning control loops can be decoupled,  and rotations are
handled by multiplication of complex exponentials. With this as a starting point, we propose and demonstrate an adaptive controller when parametric uncertainties are present. A nominal PID control design is chosen, which can guarantee closed-loop stability. This nominal design is replaced with adjustable parameters in the adaptive controller, whose parameters are updated by suitably leveraging the underlying complex state variables. Despite the state variables being complex, it is shown that a real positive definite Lyapunov function exists and that the norm of the tracking error goes to zero asymptotically. Furthermore, our results demonstrate that the proposed AC architecture outperforms a conventional PID structure in a LOE scenario. 

The contributions of the paper are an adaptive control solution for path following of a DV when parametric uncertainties are present, accommodation of input saturation, construction of a Lyapunov function despite the presence of complex state variables, proof of  stability,  and comprehensive simulation studies. To our knowledge, this is the first time that adaptive control of a Dubins Vehicle has been proposed, and therefore represents an important extension of the state of the art in path following problems when uncertainties are present.

This paper is organized as follows. Background of the DV model, trigonometric and complex state-space representations, and a compromised scenario involving a LOE are presented in Section \ref{sec:DV_background}. In Section \ref{sec:problem_formulation}, we introduce a reference model with outputs corresponding to a desired path. We then leverage the decoupling of speed and turning rate control loops to simplify the bilinear system into a linear system. In Section \ref{sec:control_design_nominal} we derive a PID controller that leads to closed loop stability of the nominal DV. In the following Section \ref{sec:control_design_adaptive}, an AC solution for the DV subject to LOE and input saturation is derived and analyzed. In Section \ref{sec:results} we present numerical experiments that validate the complex domain AC solution and make a comparison the to PID controller. We conclude the paper in Section \ref{sec:conclusion}.

\section{Dubins Vehicle: Modeling using a Complex State space}\label{sec:DV_background}
\subsection{Dubins Vehicle Model}\label{subsec:trigonometric representation}
\begin{figure}[t]
    \centering
    \includegraphics[width=1\linewidth]{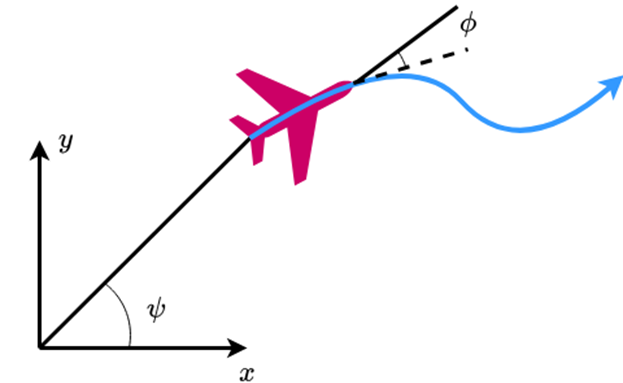}
    \caption{Fixed-wing aircraft moving along a Dubins path.}
    \label{fig:Dubins_dynamics_FBD}
\end{figure}
The equations of motion for a DV are defined as follows:
\begin{equation}
    \begin{aligned}
        \dot{x}(t) &= v(t) \cos( \psi(t) )\\
        \dot{y}(t) &= v(t) \sin( \psi(t) )\\
        \dot{v}(t) &= u_1(t) \\
        \dot{\psi}(t) &= u_2(t)
    \end{aligned}
    \label{eq:dynSYS}
\end{equation}
\noindent where $x(t), y(t) \in \mathbb{R}^2$ denote the \textit{position} of the system, $v(t) \in \mathbb{R}$ is the \textit{speed}, and $\psi(t) \in \mathbb{R}$ is the \textit{angle of the system with respect to a fixed reference frame}, and $\phi \in \mathbb{R}$ is the \textit{vehicle turning angle} (see Figure \ref{fig:Dubins_dynamics_FBD}). In \eqref{eq:dynSYS} the control inputs are $\dot{v}(t)  \in \mathbb{R}$, the \textit{speed related control} and $\dot{\psi}(t)$ is the \textit{turning control}. We also accommodate the requirement that the turning control input obey magnitude limits, i.e., require that
\begin{equation}
    \label{eq:constraint}
      \bigl|\dot{\psi}\bigr|
      \le
      \dot{\psi}_{\text{max}}, \qquad  \dot{\psi}_{\text{max}}:= \frac{g}{V_a}\tan(\phi_{c})
\end{equation}
where $g$ is \textit{gravity}, $\phi_{c} \in \mathbb{R}$ is a \textit{commanded bank angle,} $\dot{\psi}_{\text{max}}$ denotes the \textit{max turning rate} and $V_a \in \mathbb{R}^+$ is the \textit{resultant speed}.

\subsection{Complex Representations of the DV}\label{subsec:complex DV}
As seen in \eqref{eq:dynSYS} the DV model is nonlinear, making the design of control strategies challenging. To meet this challenge, we make a transformation from $\mathbb{R}^2$ to $\mathbb{C}$ as \cite{2024JGCD...47.1256L} 
\begin{equation} \label{eq:real_to_complex}
        r = x + iy\\
\end{equation}
and evaluating its derivatives. These are given by
\begin{equation}
\begin{aligned}
\label{eq:real_to_complex_derv}
    \dot{r} = \dot{x} + i\dot{y} = \sqrt{\dot{x}^2+\dot{y}^2}e^{i\psi} = V_a e^{i\psi} = v_a\\
    \dot{v}_a = (\dot{V_a} + V_a i \dot{\psi})e^{i\psi} = \bigl( \frac{\dot{V_a}}{V_a}+ i\dot{\psi} \bigr) V_ae^{i\psi} = u v_a
\end{aligned}
\end{equation}
\noindent where, $r \in \mathbb{C}$ represents \textit{position}, $v_a\in \mathbb{C}$ is the \textit{velocity} in the direction of $\psi$, $V_a \in \mathbb{R}^+$ is the \textit{resultant speed}, and $u \in \mathbb{C}$ is the \textit{control input}. We can then write the equivalent DV model dynamics in the complex domain \cite{2024JGCD...47.1256L} as
\begin{equation} 
\begin{aligned}
    \label{eq:complex_dv_model}
        \dot{r} = v_a  \\
        \dot{v}_a = \left( \frac{\dot{V}_a}{V_a} + i \dot{\psi} \right) V_a e^{i \psi} = u v_a
\end{aligned}
\end{equation}
\subsection{Parametric Uncertainty: Loss of Control Effectiveness} \label{subsec:LOE}
In real world applications, various factors like unexpected operational anomalies, actuator degradation, etc., can induce a loss of control effectiveness (LOE). One such compromised scenario occurs when turn rate control effectiveness is reduced by a factor of $ \lambda $, an unknown parameter representing \textit{control degradation}. We introduce $\lambda$ into the DV model \eqref{eq:complex_dv_model} as follows:
\begin{equation} 
\begin{aligned}
    \label{eq:compromised_system}
        \dot{r} = v_a  \\
        \dot{v}_a = \left( \frac{\dot{V}_a}{V_a} + i \lambda \dot{\psi} \right) V_a e^{i \psi}
\end{aligned}
\end{equation}
\noindent where $\lambda \in (\epsilon,1]$ and $\epsilon > 0$. It should be noted that when $\lambda=1$, there is no LOE. That is, the compromised model \eqref{eq:compromised_system} coincides with the nominal model of the DV \eqref{eq:complex_dv_model}. 
\section{Path Following Control Problem}\label{sec:problem_formulation}
\begin{figure}[t]
    \centering
    \includegraphics[width=1\linewidth]{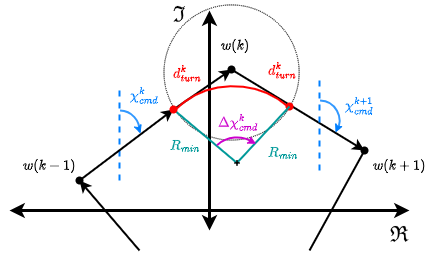}
    \caption{Path generation using complex waypoints}
    \label{fig:waypoint_path}
\end{figure}
With the DV model as in \eqref{eq:compromised_system} and the control inputs given by \eqref{eq:complex_control} we now consider the problem of choosing a control input that adapts in the presence of parametric uncertainties $\lambda$, so that a specified path $\Gamma \in \mathbb{C}$ is followed. We introduce a few definitions which we use to construct a reference model.

\subsection{Path Definition Using a Reference Model}
A path $\Gamma \in \mathbb{C}$ can be defined by a series of \textit{commanded waypoints} $\{w_c(k)\}_{k= 1,N} \in \mathbb{C}$ and \textit{commanded course angles} $\psi_{c}^k = \angle (\Delta w_c(k))$. Figure \ref{fig:waypoint_path} shows how the path between three consecutive waypoints is generated by specifying a desired turn geometry. For this purpose we introduce a \textit{minimum turning distance} $d_{turn}^k$ to define where a turn begins and ends, as
\begin{equation}
\begin{aligned}\label{eq:d_turn}
    d_{\text{turn}}^k = \frac{V_{\text{max}}^2}{g \tan(\phi_{c})} \left| \tan \left( \frac{\psi_{c}^{k+1} - \psi_{c}^k }{2} \right) \right| \\
    = R_{\text{min}} \left| \tan \left( \frac{\Delta \psi_{c}^k}{2} \right) \right|\\ 
 \end{aligned}
\end{equation}
where $R_{min} \in \mathbb{R}^+$ is the \textit{minimum turning radius}, and $V_\text{max} \in \mathbb{R}^+$ is the \textit{max speed}.  We define $\kappa_{ref} \in \mathbb{R}$ as the \textit{curvature} of the path $\Gamma \in \mathbb{C}$ such that for straight lines
\begin{equation}
    \label{eq:straight_line}
    \kappa_{\text{ref}} = 0,
\end{equation}
and for a circular path
\begin{equation} \label{eq:circular_turn}
    \kappa_{\text{ref}} = \frac{\text{sgn} (\psi_{c}^{k+1} - \psi_{c}^k)}{R_{\text{ref}}} \quad \text{,}   R_{ref} \geq R_{min}.
\end{equation}
Any path consisting of straight and turn segments of radius $R_{ref}$ can be constructed using \eqref{eq:straight_line} and \eqref{eq:circular_turn}.

For a desired path curvature $\kappa_{ref}$ and a desired speed $V_{ref} \in \mathbb{R}^+$, we can specify a desired turn rate $\dot{\psi}_{ref}$ as
\begin{equation}
    \label{eq:reference curvature}
    \dot{\psi}_{ref} = \kappa_{ref}V_{ref}.
\end{equation}
Using \eqref{eq:reference curvature} we now define the desired path that needs to be followed by a DV using a reference model that captures the desired closed-loop characteristics:
\begin{equation} 
\begin{aligned}
    \label{eq:ref_model}
        \dot{r}_{ref} = v_{ref}  \\
        \dot{v}_{ref} = \left( \frac{\dot{V}_{ref}}{V_{ref}} + i \dot{\psi}_{ref} \right) V_{ref} e^{i \psi_{ref}} = u_{ref} v_{ref}
\end{aligned}
\end{equation}
\noindent In \eqref{eq:ref_model}, $r_{ref} \in \mathbb{C}$ denotes \textit{position of the reference model}, $u_{ref} \in \mathbb{C}$ is the \textit{reference control input}, and $ v_{ref} \in \mathbb{C}$ is the \textit{reference velocity} along the direction $\psi_{ref} \in \mathbb{R}$ which is the \textit{reference model turning angle}. It should also be noted that $V_{ref}$ and $\dot{\psi}_{ref}$ can be chosen so that they are constrained as 
\begin{equation}
\begin{aligned}\label{eq:reference_constraints}
        V_\text{min} \leq V_{ref} \leq  V_\text{max} \\
        \dot{\psi}_\text{min} \leq \dot{\psi}_{ref} \leq \dot{\psi}_\text{max}
\end{aligned}
\end{equation}
in order to accommodate bounds on the turning capabilities of the DV. With \eqref{eq:ref_model} and \eqref{eq:reference_constraints}, we now have a reference model whose outputs correspond to a desired path that is a blend of straight lines and circular arcs that obey specified speed and turning rate limits.

\begin{figure}[t]
    \centering
    \includegraphics[width=1\linewidth]{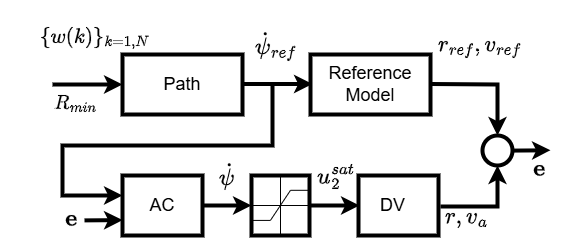}
    \caption{Block diagram of proposed path following framework.}
    \label{fig:block_diagram}
\end{figure}

\subsection{Control Transformation \& Constant Speed Simplification}\label{subsec:decomp_control}
We note from \eqref{eq:compromised_system} that the DV has two complex states given by 
\begin{equation}
    \label{eq:DV_states}
    x_d=[r, v_a]^T
\end{equation}
and that the control input $u$ 
\begin{equation}\label{eq:input_u}
    u = \frac{\dot{V}_a}{V_a} + i \lambda \dot{\psi}
\end{equation}
is a complex signal. We can however transform $u$ into real and imaginary components as follows:
\begin{equation} 
\begin{aligned}
    \label{eq:complex_control}
        u_1 = \Re(u) = \frac{\dot{V}_a}{V_a} \\
        u_2 =\Im(u) =  \dot{\psi}.
\end{aligned}
\end{equation}
That is, the real control inputs $u_1$ and $u_2$ can be viewed as inputs that separately control the vehicle speed and the turning rate, respectively. Starting with these real control inputs, we now introduce an additional assumption that $V_a$ is constant, i.e. we assume that an appropriate speed controller is in place that ensures regulation of $V_a$ around a positive constant $V_0$.
With this assumption, we now focus on a simplified dynamic model of the DV which can be obtained from the compromised DV model in \eqref{eq:compromised_system} as
\begin{equation} 
\begin{aligned}
    \label{eq:simplified_DV_system}
        \dot{r} = v_a  \\
        \dot{v}_a = i \lambda \dot{\psi} V_a e^{i \psi} = \lambda u_2 v_a .
\end{aligned}
\end{equation}
This simplification reduces the bilinear system into a linear system. 

We adopt a similar assumption and procedure for the reference model \eqref{eq:ref_model}, which has two complex states
\begin{equation}
    \label{eq:ref_model_states}
    x_{dref}=[r_{ref}, v_{ref}]^T
\end{equation}
and control input $u_{ref}$ 
\begin{equation}\label{eq:input_uref}
    u_{ref} = \frac{\dot{V}_{ref}}{V_{ref}} + i \lambda \dot{\psi}_{ref}
\end{equation}
which can be split into real and imaginary components as 
\begin{equation}
\begin{aligned}
        \label{eq:uref_decomposed}
        u_{1ref} = \Re(u_{ref}) = \frac{\dot{V}_{ref}}{V_{ref}}\\ u_{2ref} =\Im(u_{ref}) =  \dot{\psi}_{ref} \\
\end{aligned}
\end{equation}
where $u_{2ref}$ corresponds to the control input $\dot{\psi}_{ref}$. We choose $V_{ref} = V_a = V_0$,  which allows the simplification of \eqref{eq:ref_model} as
\begin{equation} 
\begin{aligned} \label{eq:simplified_ref_model}
        \dot{r}_{ref} = v_{ref}  \\
        \dot{v}_{ref} = i  \dot{\psi}_{ref} V_{ref} e^{i \psi_{ref}} =  u_{2ref} v_{ref} .
\end{aligned}
\end{equation}

\section{Nominal Control Design}\label{sec:control_design_nominal}
We now design a path following controller $u_2$ such that the DV state $x_d$ in \eqref{eq:DV_states} tracks $x_{dref}$, the state of the reference model in \eqref{eq:simplified_ref_model}, with $\dot{\psi}_{ref}$ chosen corresponding to a desired path (see Figure \ref{fig:block_diagram}). First, we define the following tracking errors
\begin{equation}
\begin{aligned}  
    \label{eq:error_defintiions}                 
    \text{Integral: } e_I = \int e_r  \ dt \\
    \text{Position: } e_r = r - r_{ref}\\
    \text{Velocity: } e_v = v_a - v_{ref}\\
\end{aligned}
\end{equation}
which capture the difference between the DV \eqref{eq:simplified_DV_system} states and reference model \eqref{eq:simplified_ref_model} states. From \eqref{eq:simplified_DV_system} \eqref{eq:simplified_ref_model}, and \eqref{eq:error_defintiions} we obtain
the error dynamics 
\begin{equation} 
\begin{aligned}
\label{eq:error_dynamics}
    \dot{e}_I = e_r\\
    \dot{e}_r = e_v  \\
    \dot{e}_v = i\lambda u_2v_a - iu_{2ref}v_{ref}. \\
\end{aligned}
\end{equation}
We will use \eqref{eq:error_dynamics} a PID control design as well as for the adaptive PID controller described later. The following lemmas are useful for the proposed nominal and adaptive controllers.

\textit{Lemma 1:} Define a complex variable $\delta \in \mathbb{C}$ as 
\begin{equation}\label{eq:ev_delta}
   \delta = i\lambda u_2v_a - iu_{2ref}v_{ref} 
\end{equation}
and $u_{\delta} \in \mathbb{C}$
\begin{equation}
    \begin{aligned}
        \label{eq:udelta}
    u_{\delta} = i \lambda u_{2ref}v_{ref} + \delta
    \end{aligned}
\end{equation}
where $v_a$ denotes the complex conjugate of $v_a$.  It can then be shown that the control input $u_2$ takes the form
\begin{equation}
        \label{eq:u2_solution}
        u_{2} =\frac{\Im\bigl(v_{a}^{*}u_{\delta}\bigr)}{\lambda V_{a}^{2}}.
\end{equation}

\noindent \textit{Proof of Lemma 1:} Substituting \eqref{eq:ev_delta} and \eqref{eq:udelta} into $\dot e_v$ leads to 
\begin{equation} \label{eq:lemma1_step1}
    u_{\delta} = i \lambda u_2 v_a.
\end{equation}

\noindent By multiplying both sides of \eqref{eq:lemma1_step1} by $v_a$ and noting that $|v_a|^2 = V_a^2$, we obtain 
\begin{equation} \label{eq:lemma1_step3}
i u_2 V_a^2  = v_a u_{\delta}  
\end{equation}
and equating imaginary parts we obtain \eqref{eq:u2_solution}. \qquad \qquad \quad $\blacksquare$

Using suitable algebraic manipulations the control solution \eqref{eq:u2_solution} can be rewritten as
\begin{equation}
        \label{eq:control_equivalence}
            u_2 =   \frac{1}{\lambda e^{i\psi}}
            \Bigl[\frac{\delta}{iV_{a}} + u_{2ref}e^{i\psi_{ref}}\Bigr].
\end{equation}

\noindent In what follows, we will use \eqref{eq:control_equivalence} for the control designs. 

\subsection{PID control for the nominal model}
We start with the nominal DV model, i.e. set $\lambda =1$ in \eqref{eq:simplified_DV_system}. We show that a PID controller can be designed for this nominal DV and that it leads to closed-loop stability. 

The starting point for the PID control design is the error dynamics in \eqref{eq:error_dynamics}. This allows a choice of $\delta$ as a PID control input in the form of 
\begin{equation} \label{eq:delta_definition}
    \delta = \mathbf{k}^\top \mathbf{e},
\end{equation}
where gains $\mathbf{k} = [k_I, k_P, k_D]^\top$ and $\mathbf{e} = [e_I, e_r, e_v]^\top$ is the state error.
Using \eqref{eq:ev_delta} and \eqref{eq:delta_definition}, the closed-loop error dynamics \eqref{eq:error_dynamics} can be written as 
\begin{equation}\label{eq:closed_loop_error}
    \dot{\mathbf{e}} =
    \begin{bmatrix}
    \dot{e}_I \\ 
    \dot{e}_r \\ 
    \dot{e}_v
    \end{bmatrix}
    =
    \underbrace{
    \begin{bmatrix}
    0 & 1 & 0 \\
    0 & 0 & 1 \\
    -k_I & -k_P & -k_D
    \end{bmatrix}
    }_{A_e}
    \begin{bmatrix}
    e_I \\ 
    e_r \\ 
    e_v
    \end{bmatrix}
    =
    A_e \mathbf{e}.
\end{equation}
One choice of the PID gains that enables $A_e$ to be a Hurwitz matrix is given by
\begin{equation} 
\begin{aligned}
    \begin{cases}
        k_I = \omega^2 a\\ 
        k_P = \omega^2 + 2 \zeta \omega a\\ 
        k_D = 2 \zeta \omega + a
    \end{cases}
\end{aligned}
\end{equation}
where $a>0$, $0\leq \zeta \leq 1$, and $\omega>0$. We note that the overall control input is given by
\begin{equation}
    \label{eq:overall_control}
    \dot{\psi} = u_2 = \frac{1}{\lambda e^{i\psi}}
            \Bigl[\frac{\mathbf{k}^\top \mathbf{e}}{iV_{a}} + u_{2ref}e^{i\psi_{ref}}\Bigr]
\end{equation}
where $u_{2ref}=\dot{\psi}_{ref}$, and the latter is given by (9). 

\section{Adaptive Control Design}\label{sec:control_design_adaptive}
In this section, we show that an AC can be designed for the compromised DV vehicle \eqref{eq:simplified_DV_system} and that it  leads to closed-loop stability and parameter learning. We first introduce an estimate {$\widehat{\theta}$ of $\frac{1}{\lambda}$} in \eqref{eq:control_equivalence} and choose an adaptive control input as
\begin{equation}
    \label{eq:lambdahat_control}
    u_2 = \frac{{\widehat{\theta}}} {e^{i\psi}}
    \left[\frac{\mathbf{k}^\top \mathbf{e}}{iV_{a}} + u_{2ref}\,e^{i\psi_{ref}}\right].
\end{equation}

Using equations \eqref{eq:error_defintiions}, \eqref{eq:lambdahat_control}, and \eqref{eq:simplified_ref_model} we obtain the velocity error dynamics
\begin{equation}
    \dot{e}_v = {\mathbf{k}}^\top \mathbf{e} + \lambda\tilde{\theta}  \Bigl({\mathbf{k}}^\top \mathbf{e} +  i u_{2\text{ref}}v_{ref}\Bigr),
\label{eq:velocity_error_with_estimate_theta}
\end{equation} 
where we define the following quantities
\begin{equation}
\label{eq:define_theta}
\hat{\theta} =\frac{1}{\hat{\lambda}},
\quad
\theta =\frac{1}{\lambda},
\quad
\tilde{\theta} =\hat{\theta} -\theta.
\end{equation}

We write the closed loop error dynamics, similar to \eqref{eq:closed_loop_error} as
\begin{equation}\label{eq:error_R}
   \dot{\mathbf{e}}=   A_e \mathbf{e} + \lambda \tilde{\theta} 
        \Bigl( \underbrace{
    \bigl( \mathbf{k}^\top \mathbf{e} + i  u_{2ref}v_{ref} \bigr)
    \begin{bmatrix}
    0 \\ 
    0 \\ 
    1
    \end{bmatrix} \Big)
    }_{R}
\end{equation}

In \eqref{eq:error_R} we note that 
$A_e, \lambda, \text{and }  \tilde{\theta}$ are real, and $R, \mathbf{e},  \text{and},  \dot{\mathbf{e}}$ are complex. Despite the presence of complex values, we leverage the fact that norms are well defined for complex variables \cite{horn2012matrix}. This in turn enables us to employ the standard Lyapunov approach leading global stability. That is, a Lyapunov function candidate is chosen as
\begin{equation} \label{eq:lyapunov_function}
    V = \bar{\mathbf{e}}^\top P \mathbf{e} + \frac{|\lambda|}{\gamma}\tilde{\theta}^2,
\end{equation}
where $\gamma>0$ is the parameter learning rate and $\bar{\mathbf{e}}$ denotes the complex conjugate $\mathbf{e}$, and $P$ is the solution of 
\begin{equation} \label{eq:P_matrix}
    A_e^T P + P A_e = -Q.
\end{equation} 
where $Q = Q^T > 0$.

It should be stressed that $\mathbf{e}$ is complex, while $\tilde{\theta}$ and $V$ are real, and that $V$ is positive definite with respect to the complex variable  $\mathbf{e}$ and the real variable $\tilde{\theta}$. Upon splitting the error $\mathbf{e}$ into real an imaginary components it is easy to see that (refer to \ref{sec:app_lyap_func})
\begin{equation} \label{eq:lyapunov_function_derivative}
    \dot{V} = 2 \mathbf{e}_{\text{real}}^\top P \dot{\mathbf{e}}_{\text{real}} 
    + 2 \mathbf{e}_{\text{imag}}^\top P \dot{\mathbf{e}}_{\text{imag}} 
    + 2 \frac{|\lambda|}{\gamma} \tilde{\theta} \dot{\tilde{\theta}}
\end{equation}
Further algebraic manipulations lead us to
\begin{equation}
    \label{eq:V_dot_plugged}
    \dot{V} = -\bar{\mathbf{e}}^\top (Q)\mathbf{e} +2\lambda\tilde{\theta}\Re\{\bar{\mathbf{e}}^\top PR\} + \frac{2|\lambda|}{\gamma}\tilde{\theta}\dot{\tilde{\theta}}.
\end{equation}
We therefore employ the adaptive law
\begin{equation} \label{eq:adaptive_law}
    \dot{\tilde{\theta}} = -\gamma\text{sign}(\lambda)\Re\{\bar{\mathbf{e}}^\top PR\}.
\end{equation}
to obtain that
\begin{equation} \label{eq:dot_V_nd}
    \dot{V} = -\bar{\mathbf{e}}^\top Q\mathbf{e} \leq 0.
\end{equation}
It is also easy to see that tracking error $\mathbf{e} \in L_2$. By applying Barbalat's lemma which states that if a uniformly continuous function has a finite $L_2$ norm then it converges to zero as $t \to \infty$ we conclude that $\mathbf{e}(t) \to 0$ as $t\rightarrow \infty$. 

In summary, we have shown in this section that for a compromised DV model with an unknown LOE $\lambda$, a steering control input $\dot{\psi}$ can be determined as
\begin{equation}
    \label{eq:adaptive_control}
    \dot{\psi} = \frac{\hat{\theta}}{e^{i\psi}}
    \left[\frac{\mathbf{k}^\top \mathbf{e}}{iV_{a}} + u_{2ref}e^{i\psi_{ref}}\right],
\end{equation}
with the corresponding adaptive law
\begin{equation} \label{eq:adaptive_law_R_spelled_out}
    \dot{\tilde{\theta}} = -\gamma\text{sign}(\lambda)\Re\{\bar{\mathbf{e}}^\top P R\}.
\end{equation}
This adaptive controller ensures that the error $\mathbf{e}$ together with $r$, $v_a$, and $\psi$ remain bounded, and that $x_d$ tracks $x_{dref}$ asymptotically. 

\subsection{Turning Rate Limits}\label{sec:control_design_saturation}
In the previous section, we derived an adaptive control law without  constraints on the heading rate $\dot{\psi}$. However, the DV is subject to turning rate limits given by
\begin{equation}\label{eq:turnmax}
\dot{\psi}_{\text{max}} = \frac{g}{V_a}\tan(\phi_{c}) = \frac{V_\text{a}}{R_\text{min}},
\end{equation}
where $\phi_{c}$ is a commanded bank angle and $R_{min}$ is the minimum turning radius of the DV. To reflect these limits, we define the saturated control input
\begin{equation}
  \label{eq:u2_sat}
  u_2^{\mathrm{sat}}(t) = \mathrm{sat}\Bigl(u_2(t), -\dot{\psi}_{\max}, \dot{\psi}_{\max}\Bigr).
\end{equation}

\subsection{Adaptive Control Design with Input Constraints}
\label{subsec:ac_control_design_saturation}
We now present the main result of this paper, which considers the control of the DV model \eqref{eq:simplified_DV_system} with the uncertainty $\lambda$, as well as input constraints \eqref{eq:constraint} so that the DV states follow those of the reference model \eqref{eq:simplified_ref_model} whose input $\dot{\psi}_{ref}$ is specified by a desired path. 

In order to accommodate the input constraints \eqref{eq:constraint}, we introduce a saturation block as shown in Figure \ref{fig:block_diagram} leading to a control input
\begin{equation} \label{eq:heading_sat}
  u_2^{\mathrm{sat}}(t)
  \;=\;
  \begin{cases}
    -\,\dot{\psi}_{\max}, & \text{if }u_2(t) < -\,\dot{\psi}_{\max}, \\
    u_2(t), & \text{if } -\dot{\psi}_{\max} \leq u_2(t) \le \dot{\psi}_{\max}, \\
    +\,\dot{\psi}_{\max}, & \text{if }u_2(t) > +\,\dot{\psi}_{\max}.
    \end{cases}
\end{equation}
This in turn can induce \emph{control clipping}  with a $\Delta_{sat}$ defined as
\begin{equation}\label{eq:u2sat_rewritten}
  \Delta_{sat} 
  = 
  u_2^{sat}-u_2
\end{equation}
which captures the difference between the desired heading-rate command $u_2$ and the actual (saturated) command $u_2^{\mathrm{sat}}$.
This in turn modifies the plant dynamics from \eqref{eq:simplified_DV_system} to the following:
\begin{equation}
\begin{aligned}
    \dot{r} = v_a,  \\
    \dot{v}_a = \bigl(i \lambda \dot{\psi}\bigr)V_ae^{i\psi}
    =
    i \lambda u_2^{\mathrm{sat}}v_a.
\end{aligned}
\label{eq:updated_plant}
\end{equation}
Next we modify the nominal reference model \eqref{eq:simplified_ref_model} by including an additional term that captures the degradation induced by $\Delta_{sat}$ such that we obtain
\begin{equation}\label{eq:degraded_reference_model}
    \begin{aligned}
        \dot{r}_{\text{ref}} &= v_{\text{ref}},\\
        \dot{v}_{\text{ref}} &= 
            \bigl(i\dot{\psi}_{ref}\bigr)v_{\text{ref}}
            +
            iV_a\hat{\lambda}\Delta_{sat}e^{i\psi}
    \end{aligned}
\end{equation}
where the second term $iV_a\widehat{\lambda}\Delta_{sat}e^{i\psi}$ quantifies the effect due to control saturation and $\widehat{\lambda}$ is an independent estimate of $\lambda$. In order to minimize control clipping, we choose the magnitude of curvature $|\kappa_{ref}|$ (and correspondingly $\dot{\psi}_{ref}$) as 
\begin{equation}
    \label{eq:max_ref_turn}
    |\kappa_{ref}| =  \frac{\lambda_{\text{min}}}{R_\text{min}}
\end{equation}
where ${\lambda_{\text{min}}}$ reflects a worst case LOE scenario. 

Following the same procedure used to obtain  \eqref{eq:velocity_error_with_estimate_theta} we obtain the velocity error dynamics (refer to Section \ref{sec:app_ac_control})
\begin{equation}\label{eq:velocity_error_saturation}
\begin{aligned}    
    \dot{e}_v 
    = 
    \mathbf{k}^\top\mathbf{e}
    +
    \lambda\tilde{\theta}\Bigl(\mathbf{k}^\top\mathbf{e} + iu_{2\text{ref}}v_{\text{ref}}\Bigr)
    -
    \tilde{\lambda}\bigl(iV_a\Delta_{sat}e^{i\psi}\bigr) \\
\text{where} \quad
\tilde{\theta} =\widehat{\theta} -\theta,
\quad
\tilde{\lambda}=\lambda-\widehat{\lambda}.
\end{aligned}
\end{equation} 
The additional term $-\tilde{\lambda}iV_a\Delta_{sat}e^{i\psi}$
reflects the effect of saturation on the velocity error. Following a similar procedure used to get \eqref{eq:error_R}, we obtain the following closed-loop error dynamics 
\begin{equation}\label{eq:error_S}
    \dot{\mathbf{e}}
    =
    A_{e}\mathbf{e}
    +
    \lambda\tilde{\theta}R
    -
    -
    \tilde{\lambda} 
    \underbrace{iV_a\Delta_{sat} e^{i\psi}
    \begin{bmatrix}
        0 \\[4pt]
        0 \\[4pt]
        1
    \end{bmatrix}}_{S}.
\end{equation}

In \eqref{eq:error_S} we note that $A_e, \lambda, \text{and},  \tilde{\theta}, \tilde{\lambda}$ are real, and $R, \mathbf{e},  \text{and},  \dot{\mathbf{e}}$ are complex. We use a Lyapunov approach similar to \eqref{eq:lyapunov_function}, but we now extend the Lyapunov function to include an additional term involving $\tilde{\lambda}$ as follows
\begin{equation}\label{eq:lyapunov_function_saturation}
  V 
  =
  \bar{\mathbf{e}}^\top P\mathbf{e}
  +
  \frac{|\lambda|}{\gamma_{\theta}}\tilde{\theta}^2
  +
  \frac{1}{\gamma_{\lambda}}\tilde{\lambda}^2,
\end{equation}
where the matrix $P$ is positive definite, and $\gamma_{\theta}, \gamma_{\lambda}>0$ are adaptation gains (learning rates). It is important to note that $\mathbf{e}$ is complex and $\tilde{\theta}$, $\tilde{\lambda}$ and $V$ are real. $V$ is positive definite with respect to the complex variable  $\mathbf{e}$ and the real variables $\tilde{\theta}$, $\tilde{\lambda}$. 
Upon splitting the error $\mathbf{e}$ into real an imaginary components it is easy to see that
\begin{equation}
    \dot{V} 
    =
    2 \mathbf{e}_{\text{real}}^\top P \dot{\mathbf{e}}_{\text{real}} 
    + 2 \mathbf{e}_{\text{imag}}^\top P \dot{\mathbf{e}}_{\text{imag}} 
    + 2 \frac{|\lambda|}{\gamma} \tilde{\theta} \dot{\tilde{\theta}}
    \label{eq:lyapunov_function_derivative_saturation}
\end{equation}
Further algebraic manipulations lead us to
\begin{equation}
\begin{aligned}
    \label{eq:V_dot_plugged_sat}
    \dot{V} = -\bar{\mathbf{e}}^\top Q \mathbf{e} +2\lambda\tilde{\theta}\Re\{\bar{\mathbf{e}}^\top PR\}\\- 2\tilde{\lambda} \Re\{\bar{\mathbf{e}}^\top P S\} + \frac{2|\lambda|}{\gamma}\tilde{\theta}\dot{\tilde{\theta}} + \frac{2}{\gamma}\tilde{\lambda}\dot{\tilde{\lambda}}.
    \end{aligned}
\end{equation}
We therefore employ the following adaptive laws
\begin{equation}
  \dot{\tilde{\theta}} 
  = 
  -\gamma_{\theta}\mathrm{sign}\bigl(\lambda\bigr)\Re\Bigl\{\bar{\mathbf{e}}^\top PR\Bigr\},
  \quad
  \dot{\tilde{\lambda}}
  =
  \gamma_{\lambda}\Re\Bigl\{\bar{\mathbf{e}}^\top PS\Bigr\},
  \label{eq:adaptive_laws}
\end{equation}
to obtain that
\begin{equation} \label{eq:dot_V_nd_2}
    \dot{V} = -\bar{\mathbf{e}}^\top Q\mathbf{e} \leq 0.
\end{equation}
and ensure $\dot{V}\le0$.  Equations \eqref{eq:lyapunov_function_saturation} and \eqref{eq:dot_V_nd_2} imply that all solutions of the adaptive system defined by equations \eqref{eq:error_S} and \eqref{eq:adaptive_laws} are globally bounded. It is also easy to see that  $\mathbf{e} \in L_2$. By applying Barbalat's lemma which states that if a uniformly continuous function has a finite $L_2$ norm, we can conclude that $\mathbf{e}(t) \to 0$ as $t\rightarrow \infty$. In summary, this establishes that the error $\mathbf{e}$ together with $r$, $v_a$, and $\psi$ remain bounded, and that $x_d$ tracks $x_{dref}$ asymptotically.  It should be noted that (i) the adaptive laws in \eqref{eq:adaptive_laws} adjust the control parameters $\widehat{\theta}$ and $\widehat{\lambda}$ in a way that compensates for both the parametric uncertainty and the extra “degradation” term caused by saturation; (ii) the derivative of $V$ is computable even though $\mathbf{e} \in \mathbb{C}$ due to the fact that $V$ can be expressed completely in terms of the real variables $\mathbf{e}_\text{real}$ and $\mathbf{e}_\text{imag}$.

\section{Numerical Experiments}\label{sec:results}
\begin{figure}[!h]
    \centering
    \includegraphics[width=0.5\linewidth]{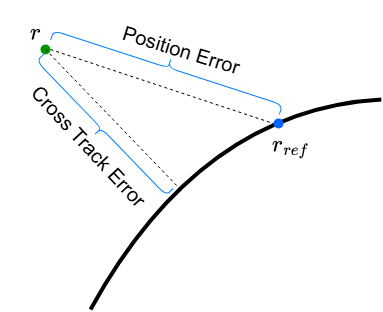}
    \caption{Illustration depicting the difference between position error and cross track error. $\lambda$.}
    \label{fig:error_dif}
\end{figure}

\begin{figure}[!h]
    \centering
    \includegraphics[width=1\linewidth]{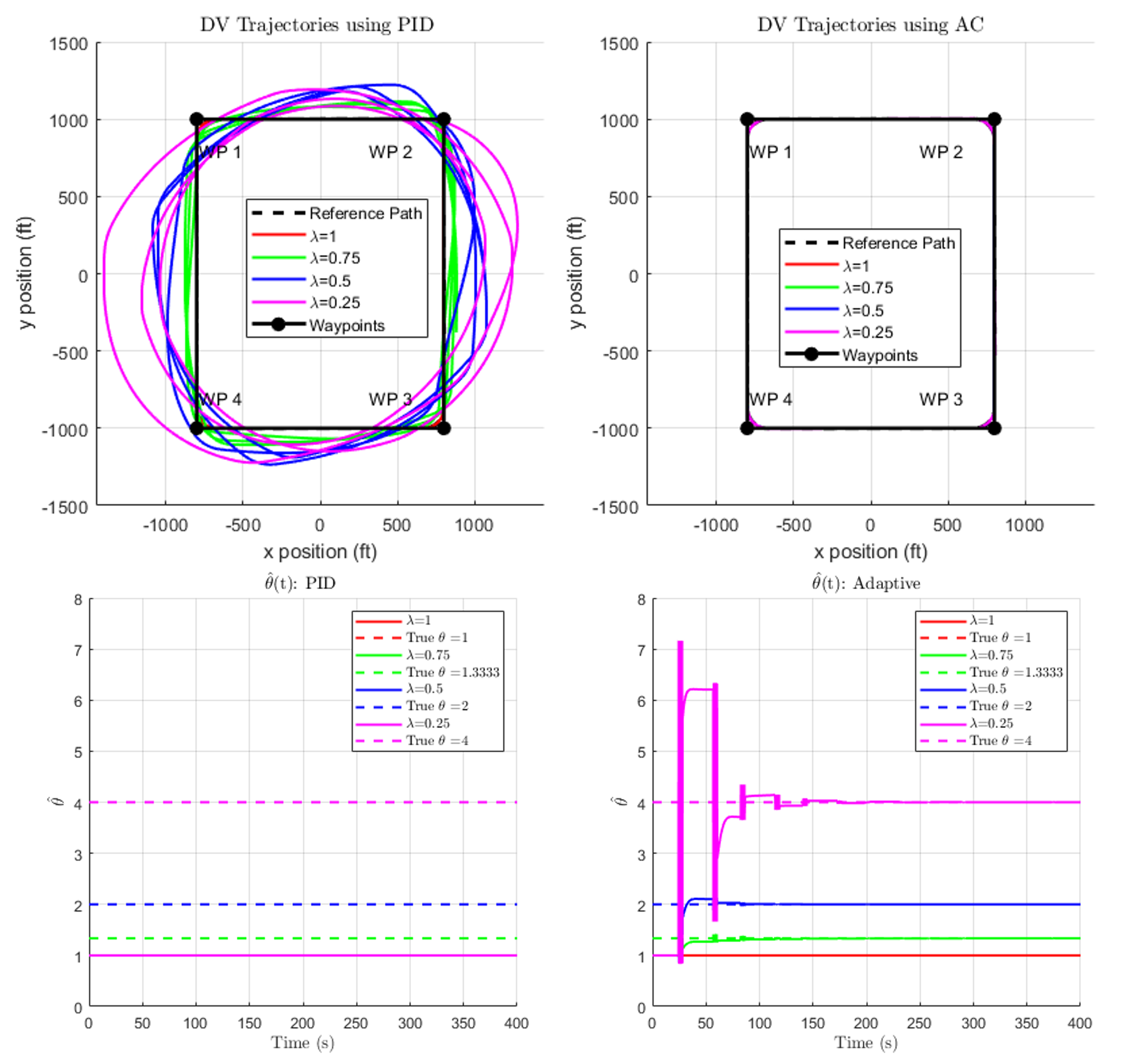}
    \caption{Comparison between vehicle trajectories with a PID control design (left) vs AC design (right). First Row: Dubin Vehicle trajectories for values of $\lambda = [1,0.75,0.5,0.25]$. Second Row: Parameter estimates for $\hat{\theta}$ (solid lines) and true values (dashed lines).} 
    \label{fig:nosat_trajectories}
\end{figure}

\begin{figure}[!h]
    \centering

    \includegraphics[width=1\linewidth]{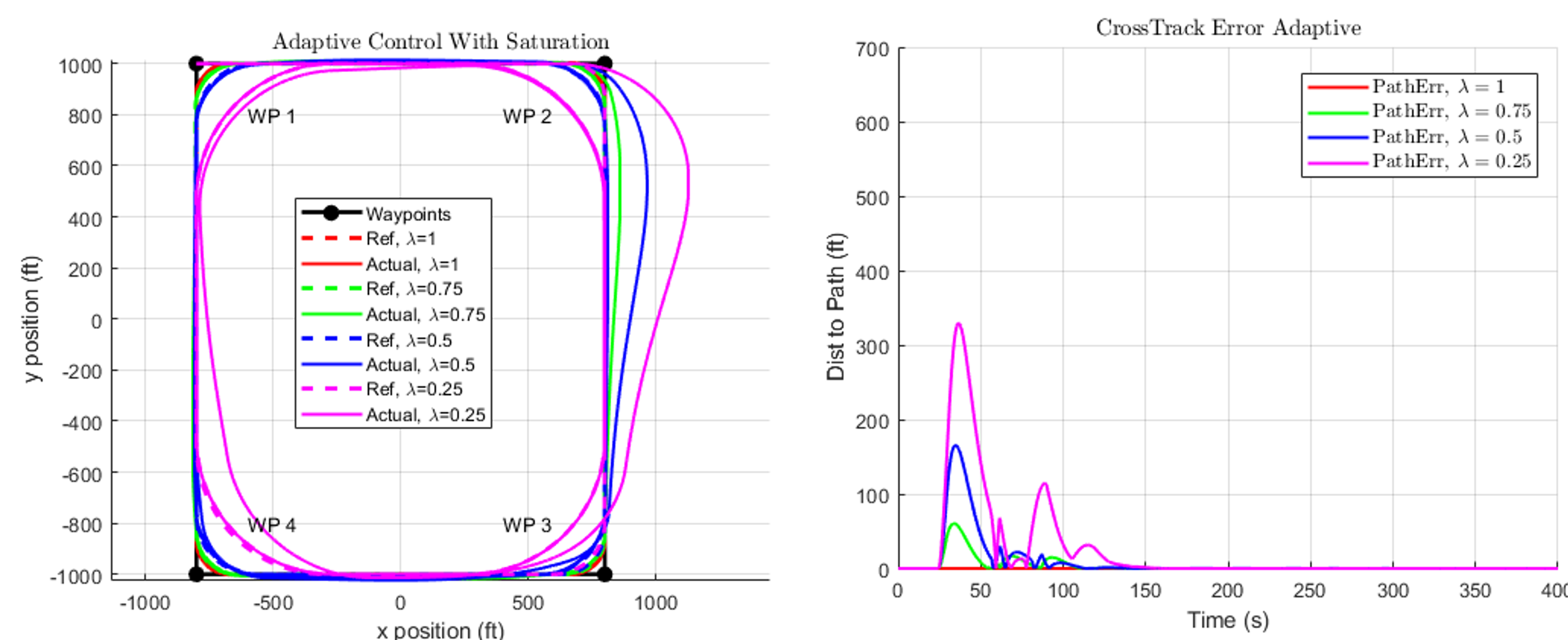}
    \caption{Dubin Vehicle trajectories under an Adaptive Control (AC) law with input saturation (left), corresponding cross-track error (right). The rectangular reference path is shown with waypoints (black dots), and each colored trajectory/error/estimate corresponds to one of the four LOE scenarios $\lambda = \{1, 0.75, 0.5, 0.25\}$.}
    \label{fig:saturation_trajectories}
\end{figure}

This section evaluates DV path following under parametric uncertainty and input constraints using the PID \eqref{eq:overall_control} and AC \eqref{eq:adaptive_control}. In our simulation, $V_a = V_{ref}$ is set to 60 ft/s, $\lambda_{min}$ is 0.25, the design parameter $a$ is 0.1, $\zeta$ is 0.8, the  $\omega$ is 0.1 rad/s, $g$ is 32.2 ft/s$^2$, $\phi_{c}$ is $45^\circ$, $\dot{\psi}_{\text{max}}$ is $30.75^\circ$ (deg/s), and $R_{min}$ is 134.2 ft. We use a rectangular path generated using \eqref{eq:simplified_ref_model}, and test four LOE scenarios: $\lambda = 1, 0.75, 0.5, 0.25$. The DV starts at the first waypoint, and travels clockwise along the waypoints for a total time $t = 400$ seconds. Using the above scenario we compare the performance of PID and AC controllers using the tracking errors \eqref{eq:error_defintiions} and a cross track error (see Figure \ref{fig:error_dif} for the distinction between position and cross-track errors). Figures \ref{fig:nosat_trajectories} and \ref{fig:saturation_trajectories} show DV trajectories (magenta, cyan, green, pink) corresponding to the four LOE scenarios.
Figure \ref{fig:nosat_trajectories} shows that with the PID controller, tracking performance begins to degrade from $\lambda = 0.75$ onward, causing significant overshoots and tracking errors. In contrast, over time the AC estimates $\hat{\theta}$, (solid) which converge to the true values (dashed), yielding better path following. 

\begin{figure}[!h]
    \centering
    \includegraphics[width=1\linewidth]{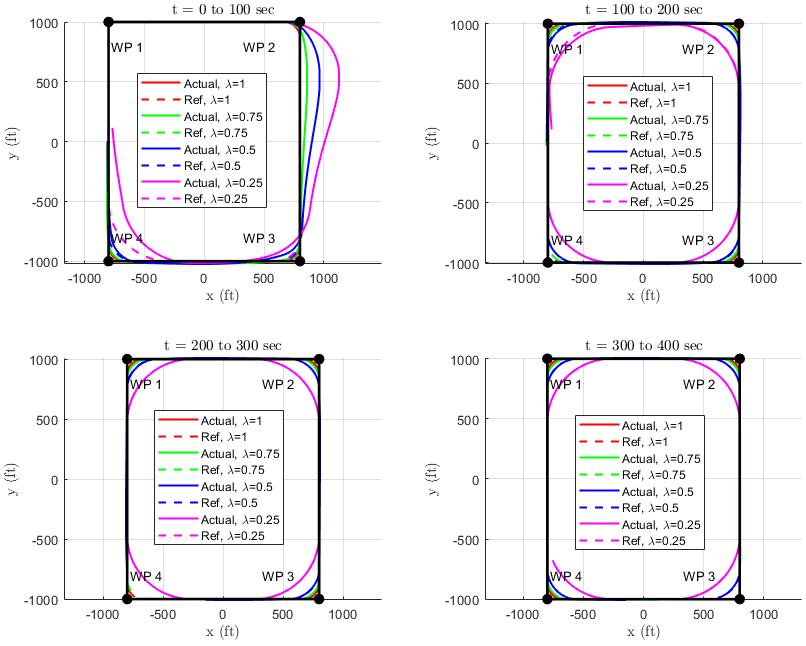}
    \caption{Trajectory-tracking results for a vehicle moving between four waypoints in a clockwise direction, starting at WP1. Each subplot highlights a different time interval (top-left) $t=0$ to $100$ s, (top-right) $t=100$ to $200$ s, (bottom-left) $t=200$ to $300$ s, and (bottom-right) $t=300$ to $400$ s. Black markers denote the waypoints, dashed lines show the reference path, and the solid lines represent the actual trajectories under four LOE scenarios $\lambda = \{1, 0.75, 0.5, 0.25\}$.}
    \label{fig:snapshot}
\end{figure}

As shown in Figure \ref{fig:saturation_trajectories}, the Adaptive Controller (AC) follows the prescribed path. The left subplot shows that for each loss-of-effectiveness (LOE) scenario, the actual Dubin Vehicle (DV) path (colored lines) closely follows to the rectangular path. The AC successfully reduces the cross track error, which is plotted in the right subplot. Table \ref{tab:all-lambda} quantifies these improvements, with AC reducing velocity, cross-track, and position errors by a orders of magnitude. 
\begin{figure}[!h]
    \centering
    \includegraphics[width=1\linewidth]{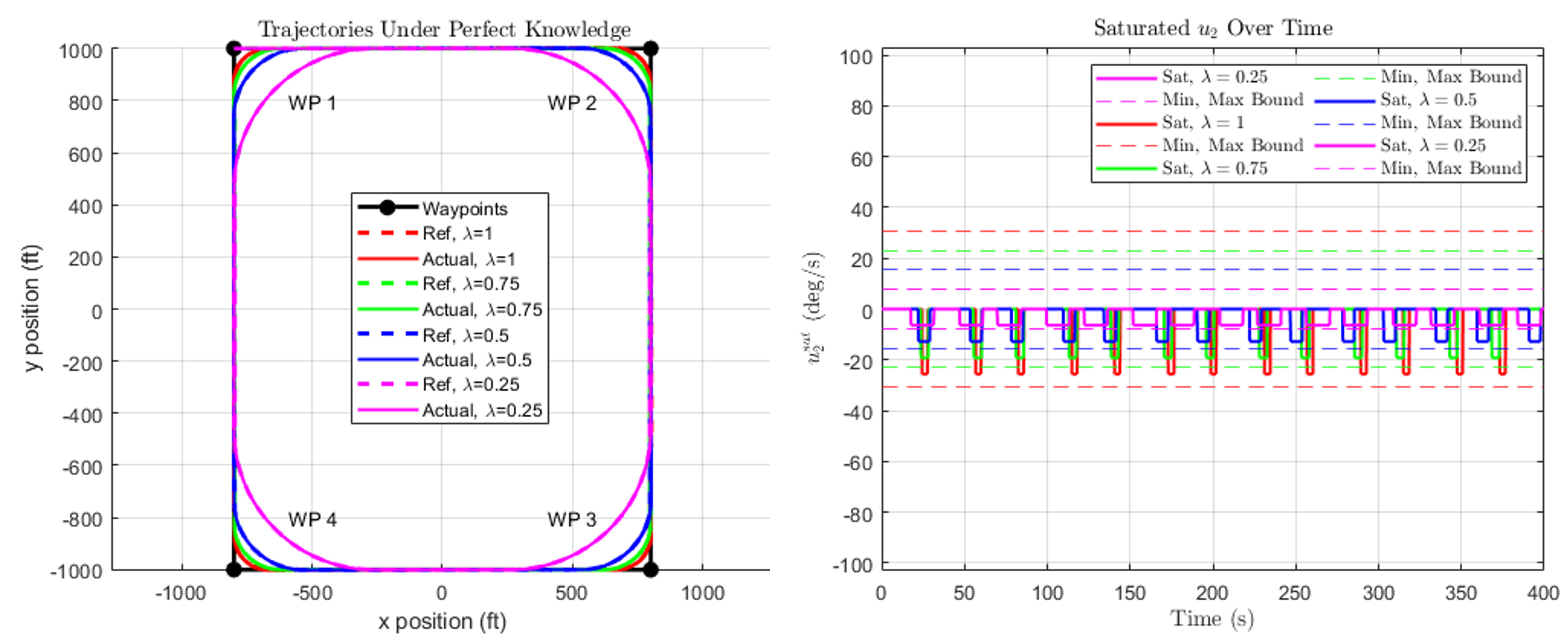}
    \caption{Dubin Vehicle trajectories (left) and control input (right) with perfect knowledge of $\lambda$. The rectangular reference path is shown with waypoints (black dots), and each colored trajectory/error/estimate corresponds to one of the four LOE scenarios $\lambda = \{1, 0.75, 0.5, 0.25\}$.}
    \label{fig:perfect}
\end{figure}
In Figure \ref{fig:perfect} we show the simulation results under the assumption of perfect knowledge of the LOE parameter $\lambda$. We will then use this plot to make a comparison with the steady state adaptive control solution. In Figure \ref{fig:saturation_trajectories} it is hard to distinguish the transient and steady state behavior so we make Figure \ref{fig:snapshot}, which plots snapshots of the trajectory over intervals of 100 seconds. 

Figure \ref{fig:snapshot} highlights the transient and steady-state behaviors. The top-left plot shows the transient behavior is most noticeable in the first few turns while the other plots show the steady-state behavior. Comparing the steady-state behavior of for each $\lambda = \{1, 0.75, 0.5, 0.25\}$ to the perfect knowledge simulation results shown in Figure \ref{fig:perfect} we see that they match very closely. 

\newpage
\begin{figure}[!h]
    \centering

    \includegraphics[width=1\linewidth]{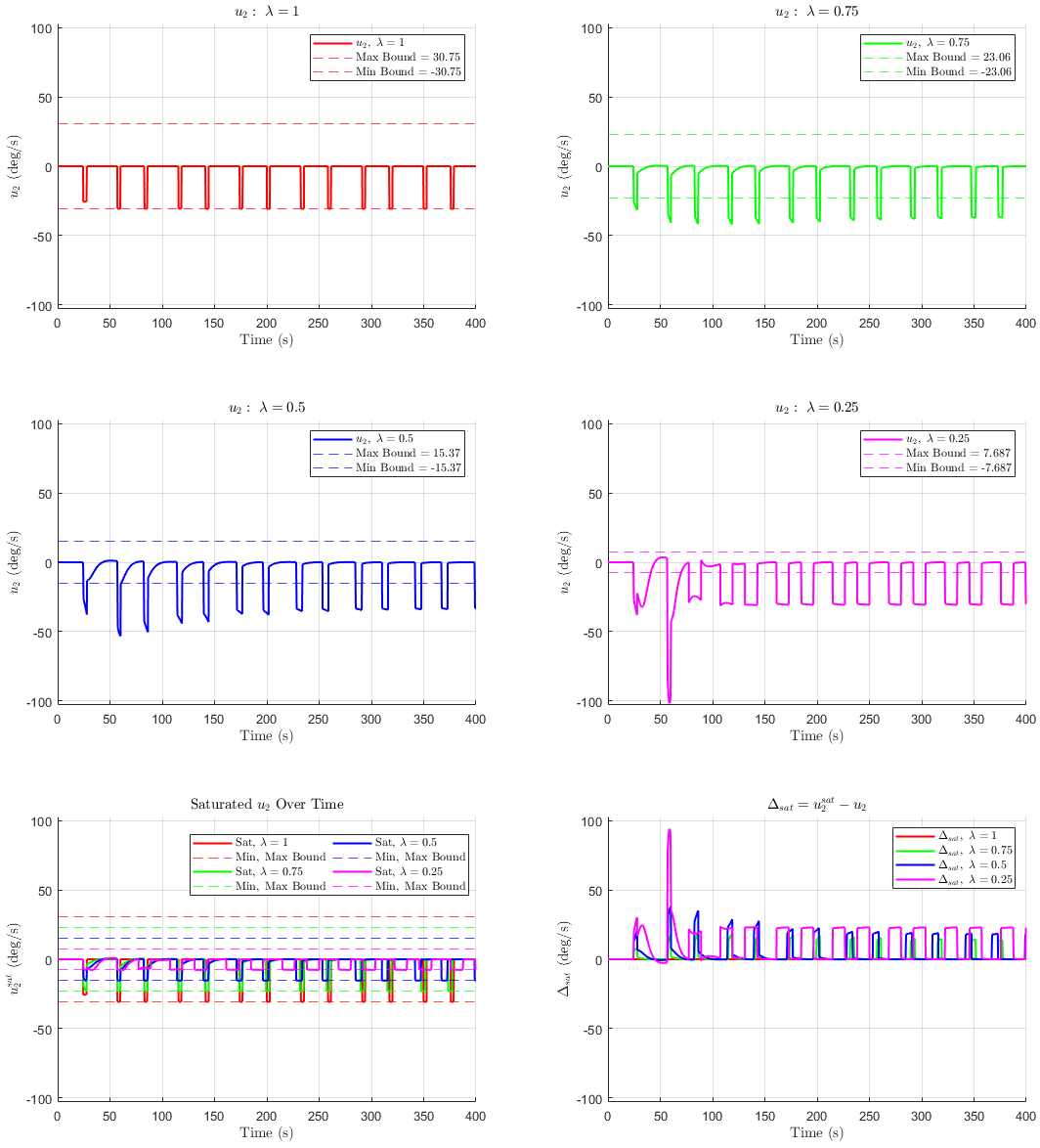}
    \caption{The control $u_2$ (in degrees) in the four LOE scenarios of $\lambda = \{1, 0.75, 0.5, 0.25\}$. Each subplot shows the actual control (solid line), its maximum and minimum bounds (dashed lines), and the reference value (dash-dotted line). The bottom-left subplot shows $u_2^{sat}$ (in degrees) over time. The bottom-right subplot depicts the control-input difference, $\Delta_{u_2} = u_2^{sat} - u_2$.}
    \label{fig:control}
\end{figure}
By inspecting the first four subplots in Figure \ref{fig:control}, it is apparent that saturation occurs in the LOE scenarios where $\lambda = \{0.75, 0.5, 0.25\}$. The bottom-left subplot shows $u_2^{sat}$ $\dot{\psi}$ stays within the control bounds. The bottom right plot shows how the more severe the LOE (the smaller the value of $\lambda$) the greater the control clipping $\Delta_{sat}$. 

\begin{figure}[!h]
    \centering
    \includegraphics[width=1\linewidth]{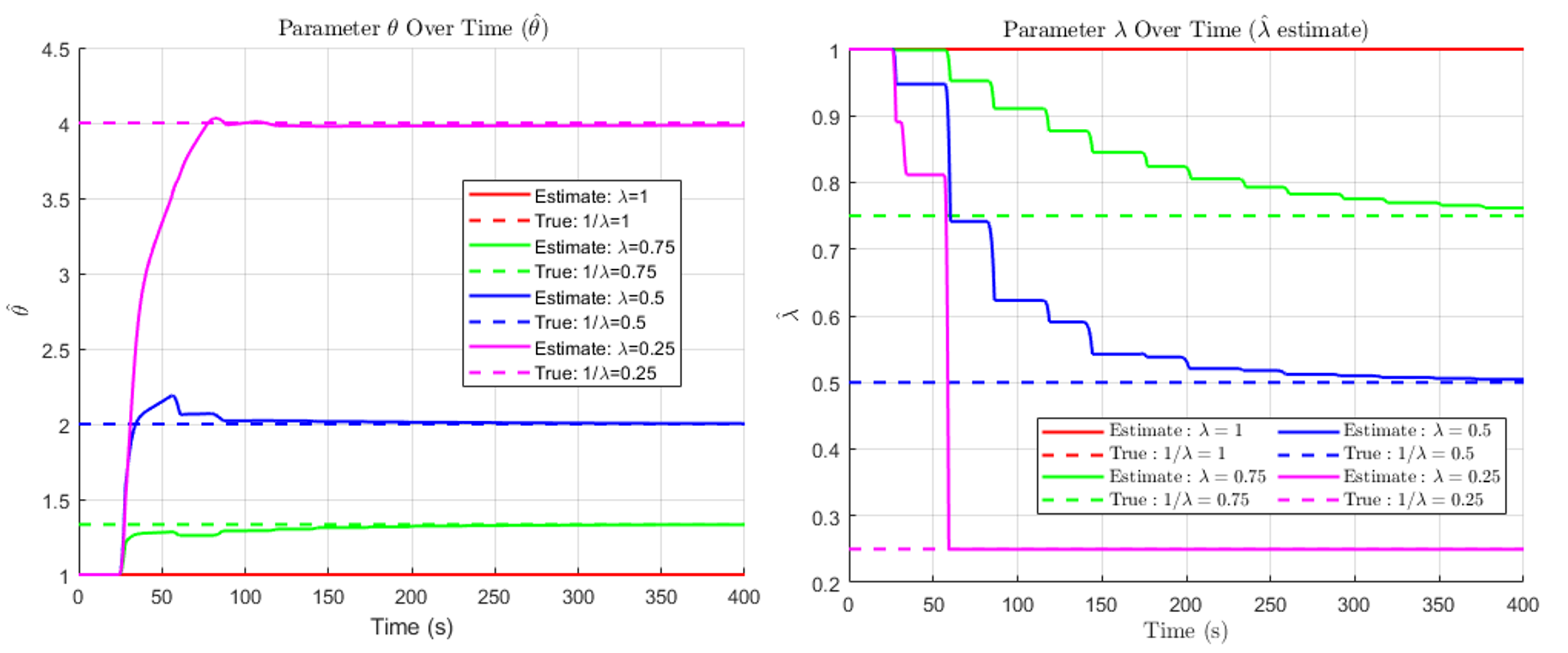}
    \caption{Time evolution of the estimated parameters for different LOE scenarios. The left subplot shows 
    $\hat{\theta}$ (solid lines) 
    alongside their true values (dashed lines), while the right subplot presents 
    $\hat{\lambda}$ (solid lines) with corresponding true values (dashed lines). 
    Both subplots highlight the parameter adaptation process for each LOE scenario $\lambda = \{1, 0.75, 0.5, 0.25\}$.}
    \label{fig:parameters}
\end{figure}

Figure \ref{fig:parameters} illustrates how the parameters $\hat{\theta}$ and $\hat{\lambda}$ evolve over time such that  $\hat{\theta}\rightarrow \theta$ and $\hat{\lambda}\rightarrow \lambda$. Table \ref{tab:all-lambda} shows the mean and standard deviation of the following error metrics: velocity, heading, position and cross-track errors for each LOE scenario $\lambda = \{1, 0.75, 0.5, 0.25\}$.

\begin{table}[h]
\caption{Error Metrics for PID and Adaptive Controllers under varying levels of control effectiveness $\lambda$ values.}
\label{tab:all-lambda}
\centering
\begin{tabular}{c l c c}
\hline
\textbf{$\lambda$} & \textbf{Metric}         & \textbf{PID}           & \textbf{Adaptive}      \\
\hline
1    & Velocity Error           & $0.000 \pm 0.000$      & $0.000 \pm 0.000$ \\
     & Heading Err (deg)   & $0.003 \pm 0.008$      & $0.003 \pm 0.008$ \\
     & Pos Err (ft)        & $0.000 \pm 0.000$      & $0.000 \pm 0.000$ \\
     & CrossTrack Err (ft) & $0.000 \pm 0.000$      & $0.000 \pm 0.000$ \\
\hline
0.75 & Velocity Error           & $5.845 \pm 5.622$      & $0.000 \pm 0.001$ \\
     & Heading Err (deg)   & $5.590 \pm 5.386$      & $0.003 \pm 0.008$ \\
     & Pos Err (ft)        & $89.438 \pm 45.118$    & $0.001 \pm 0.003$ \\
     & CrossTrack Err (ft) & $47.173 \pm 29.028$    & $0.001 \pm 0.003$ \\
\hline
0.5  & Velocity Error           & $21.176 \pm 16.433$    & $0.000 \pm 0.001$ \\
     & Heading Err (deg)   & $20.564 \pm 16.197$    & $0.003 \pm 0.008$ \\
     & Pos Err (ft)        & $374.695 \pm 187.831$  & $0.001 \pm 0.005$ \\
     & CrossTrack Err (ft) & $122.040 \pm 78.834$   & $0.001 \pm 0.005$ \\
\hline
0.25 & Velocity Error           & $47.360 \pm 26.036$    & $0.001 \pm 0.003$ \\
     & Heading Err (deg)   & $48.003 \pm 27.889$    & $0.003 \pm 0.009$ \\
     & Pos Err (ft)        & $871.815 \pm 359.424$  & $0.004 \pm 0.010$ \\
     & CrossTrack Err (ft) & $170.057 \pm 152.174$  & $0.004 \pm 0.010$ \\
\hline
\end{tabular}
\end{table}

\section{Conclusion}\label{sec:conclusion}
Dubins vehicle (DV) represents a canonical model of a vehicle that can be used for designing control methods for path-following and waypoint guidance problems in several applications. In this paper, we leverage the complex representation of the DV and present the full problem formulation under parametric uncertainties. Using a Lyapunov-based stability proof and explicit handling of input saturation, we show that the proposed adaptive controller guarantees closed-loop boundedness and ensures that the DV follows the desired path even when subjected to a LOE. While the focus of this paper is only when the parametric uncertainty is in the form of a control loss of effectiveness, extensions to other uncertainties can be carried out in a straight forward manner. 

The paper assumes that a speed control loop can be designed, thereby overcoming the challenges introduced by the underlying bilinearity. Future research will address a simultaneous design of both the control of the turning radius as well as the vehicle speed. In addition to input constraints, state constraints such as no fly zones will also be addressed. In this study, we focused on the 2D DV, and we plan to extend our work to the 3D DV in future research.
\bibliographystyle{ieeetr}
\bibliography{references}

\appendix
\subsection{Lyapunov Function Analysis}
\label{sec:app_lyap_func}
Taking the closed loop error dynamics Eq. \eqref{eq:error_R}
\begin{equation}
\begin{aligned}
\dot{\mathbf{e}} &=
\begin{bmatrix}
\dot{e}_I \\
\dot{e}_r \\
\dot{e}_v
\end{bmatrix}
=
A_e \mathbf{e}
+ \lambda \tilde{\theta} \underbrace{\left( \mathbf{k}^T \mathbf{e} + i u_{2\text{ref}} v_{\text{ref}} \right)}_{R}
\begin{bmatrix}
0 \\
0 \\
1
\end{bmatrix}
\end{aligned}
\label{eq:closed_loop_error_dynamics_35}
\end{equation}
where $A_e$ \( \in \mathbb{R}^{3\times3}\) is a Hurwitz (stable) real matrix, $\lambda, \tilde{\theta} \in \mathbb{R}$, 
$e_I, e_r, e_v, \dot{e}_I, \dot{e}_r, \dot{e}_v$ are complex $\in \mathbb{C}$ and $R \in \mathbb{C}^3$. It is possible to split \eqref{eq:closed_loop_error_dynamics_35} into real and imaginary components as follows
\begin{equation}
\begin{aligned}
\dot{\mathbf{e}}_{\text{real}} &= A_e \Re(\mathbf{e}) + \lambda \tilde{\theta} \Re(R) \\
\dot{\mathbf{e}}_{\text{imag}} &= A_e \Im(\mathbf{e}) + \lambda \tilde{\theta} \Im(R). \\
\end{aligned}
\label{eq:combined_error_dynamics_35_real_imag_split}
\end{equation}
We can obtain \eqref{eq:closed_loop_error_dynamics_35} using \eqref{eq:combined_error_dynamics_35_real_imag_split} and performing the following
\begin{equation}
\begin{aligned}
\dot{\mathbf{e}} &= \dot{\mathbf{e}}_{\text{real}} + i \dot{\mathbf{e}}_{\text{imag}} \\
\dot{\mathbf{e}} &= A_e \Re(\mathbf{e}) + \lambda \tilde{\theta} \Re(R) + i \left( A_e \Im(\mathbf{e}) + \lambda \tilde{\theta} \Im(R) \right) \\
\dot{\mathbf{e}} &= A_e \left( \underbrace{\Re(\mathbf{e}) + i \Im(\mathbf{e})}_{\mathbf{e}} \right)
+ \lambda \tilde{\theta} \left( \underbrace{\Re(R) + i \Im(R)}_{R} \right) \\
\dot{\mathbf{e}} &= A_e \mathbf{e} + \lambda \tilde{\theta} R
\end{aligned}
\label{eq:combined_error_dynamics}
\end{equation}

The choice of Lyapunov Function Eq. \eqref{eq:lyapunov_function}
\begin{itemize}
  \item $\mathbf{e} \in \mathbb{C}^3$
    \item $P \in \mathbb{R}^{3\times3}$ is a symmetric positive-definite matrix chosen to satisfy
    \begin{equation}\label{eq:Lyap_eq}
    A_e^\top P + P A_e = -Q,\quad Q = Q^\top > 0,
    \end{equation}
  \item $\lambda, \tilde{\theta} \in \mathbb{R}$ 
  \item $\gamma > 0$
  \item $\bar{\mathbf{e}}$: complex conjugate of $\mathbf{e}$, so $\bar{\mathbf{e}}^\top$ is its conjugate transpose
  \item $\mathbf{e}_{\text{real}} = \Re(\mathbf{e})$, \quad $\mathbf{e}_{\text{imag}} = \Im(\mathbf{e})$ which are both  $\in \mathbb{R}^{3\times 1}$
  \item $R_{\text{real}} = \Re(R)$, \quad $R_{\text{imag}} = \Im(R)$ which are both  $\in \mathbb{R}^{3\times 1}$
\end{itemize}

It is possible to split $\bar{\mathbf{e}}$ and ${\mathbf{e}}^\top$ into real and imaginary parts as follows
\begin{equation}
\begin{aligned}    
    \label{eq:split_errors_imag_real}
    \bar{\mathbf{e}} = \mathbf{e}_{\text{real}} - i \mathbf{e}_{\text{imag}}\\
    {\mathbf{e}} = \mathbf{e}_{\text{real}} + i \mathbf{e}_{\text{imag}}\\
    \end{aligned}
\end{equation}
which we can then substitute \eqref{eq:split_errors_imag_real} into \eqref{eq:lyapunov_function}
\begin{equation}
\begin{aligned}
V &= \bar{\mathbf{e}}^\top P \mathbf{e} + \frac{|\lambda|}{\gamma} \tilde{\theta}^2 \\
  &= (\mathbf{e}_{\text{real}} - i \mathbf{e}_{\text{imag}})^\top P (\mathbf{e}_{\text{real}} + i \mathbf{e}_{\text{imag}}) + \frac{|\lambda|}{\gamma} \tilde{\theta}^2 \\
  &= (\mathbf{e}_{\text{real}}^\top - i \mathbf{e}_{\text{imag}}^\top)(P \mathbf{e}_{\text{real}} + i P \mathbf{e}_{\text{imag}}) + \frac{|\lambda|}{\gamma} \tilde{\theta}^2 \\
  &= \mathbf{e}_{\text{real}}^\top P \mathbf{e}_{\text{real}}  + \mathbf{e}_{\text{imag}}^\top P \mathbf{e}_{\text{imag}} + \frac{|\lambda|}{\gamma} \tilde{\theta}^2
\end{aligned}
\label{eq:method1_split_real_imag}
\end{equation}
where P is real and symmetric
\begin{equation}
\begin{aligned}
\label{eq:scalar_transpose}
\mathbf{e}_{\text{real}}^\top P \mathbf{e}_{\text{imag}} &= (\mathbf{e}_{\text{imag}}^\top P \mathbf{e}_{\text{real}})^\top = \mathbf{e}_{\text{imag}}^\top P \mathbf{e}_{\text{real}} \end{aligned}
\end{equation}
The last equality holds since $\mathbf{e}_{\text{imag}}^\top P \mathbf{e}_{\text{real}}$ is a scalar. {\bf This enables us to take the derivative of $V$ only with respect to real-valued variables. This is the main reason why \eqref{eq:lyapunov_function} is valid.}

\begin{equation}
\begin{aligned}
V &= \mathbf{e}_{\text{real}}^\top P \mathbf{e}_{\text{real}} + \mathbf{e}_{\text{imag}}^\top P \mathbf{e}_{\text{imag}} \\ +& i \underbrace{\left( \mathbf{e}_{\text{real}}^\top P \mathbf{e}_{\text{imag}} - \mathbf{e}_{\text{imag}}^\top P \mathbf{e}_{\text{real}} \right)}_{0} + \frac{|\lambda|}{\gamma} \tilde{\theta}^2 \\
V &= \underbrace{\mathbf{e}_{\text{real}}^\top P \mathbf{e}_{\text{real}} + \mathbf{e}_{\text{imag}}^\top P \mathbf{e}_{\text{imag}}}_{\bar{\mathbf{e}}^\top P \mathbf{e}} + \frac{|\lambda|}{\gamma} \tilde{\theta}^2 \\
\end{aligned}
\label{eq:real_symmetric_P_simplification}
\end{equation}
so $V \in \mathbb{R}$ because the imaginary term is equal to 0. Next we compute $\dot{V}$
\begin{equation}
\begin{aligned}
\dot{V} &= \left( \frac{\partial V}{\partial \mathbf{e}_{\text{real}}} \right)^\top \dot{\mathbf{e}}_{\text{real}} 
+ \left( \frac{\partial V}{\partial \mathbf{e}_{\text{imag}}} \right)^\top \dot{\mathbf{e}}_{\text{imag}} 
+ \frac{\partial V}{\partial \tilde{\theta}} \dot{\tilde{\theta}} \\
&= 2 \mathbf{e}_{\text{real}}^\top P \dot{\mathbf{e}}_{\text{real}} 
+ 2 \mathbf{e}_{\text{imag}}^\top P \dot{\mathbf{e}}_{\text{imag}} 
+ 2 \frac{|\lambda|}{\gamma} \tilde{\theta} \dot{\tilde{\theta}}
\end{aligned}
\label{eq:V_dot_derivative}
\end{equation}
where the partial derivatives are all with respect to real valued variables, ie. we are no longer taking the partial derivative of $V$ with respect to complex variable $\mathbf{e}$. 

Next we substitute in \eqref{eq:combined_error_dynamics_35_real_imag_split} into \eqref{eq:V_dot_derivative}
\begin{equation}
\begin{aligned}
\dot{V} &= 2 \mathbf{e}_{\text{real}}^\top P \left( A_e \mathbf{e}_{\text{real}} + \lambda \tilde{\theta} R_{\text{real}} \right) \\ 
&+ 2 \mathbf{e}_{\text{imag}}^\top P \left( A_e \mathbf{e}_{\text{imag}} + \lambda \tilde{\theta} R_{\text{imag}} \right)
+ 2 \frac{|\lambda|}{\gamma} \tilde{\theta} \dot{\tilde{\theta}} \\
&= 2 \mathbf{e}_{\text{real}}^\top P A_e \mathbf{e}_{\text{real}} + 2 \mathbf{e}_{\text{imag}}^\top P A_e \mathbf{e}_{\text{imag}} \\ & 
+ 2 \lambda \tilde{\theta} \left( \mathbf{e}_{\text{real}}^\top P R_{\text{real}} + \mathbf{e}_{\text{imag}}^\top P R_{\text{imag}} \right)
+ 2 \frac{|\lambda|}{\gamma} \tilde{\theta} \dot{\tilde{\theta}}
\label{eq:V_dot_expanded_1}
\end{aligned}
\end{equation}
where we can look at the first term in \eqref{eq:V_dot_expanded_1}

\begin{equation}
\begin{aligned}
{2 \mathbf{e}_{\text{real}}^\top P A_e \mathbf{e}_{\text{real}}}
&= \mathbf{e}_{\text{real}}^\top A_e^\top P \mathbf{e}_{\text{real}} + \mathbf{e}_{\text{real}}^\top P A_e \mathbf{e}_{\text{real}} 
\\&= \mathbf{e}_{\text{real}}^\top (A_e^\top P + P A_e) \mathbf{e}_{\text{real}} 
\label{eq:V_dot_expanded_comp1}
\end{aligned}
\end{equation}
and the second term in \eqref{eq:V_dot_expanded_1}
\begin{equation}
\begin{aligned}
{2 \mathbf{e}_{\text{imag}}^\top P A_e \mathbf{e}_{\text{imag}}}
&= \mathbf{e}_{\text{imag}}^\top A_e^\top P \mathbf{e}_{\text{imag}} + \mathbf{e}_{\text{imag}}^\top P A_e \mathbf{e}_{\text{imag}} 
\\&= \mathbf{e}_{\text{imag}}^\top (A_e^\top P + P A_e) \mathbf{e}_{\text{imag}}.
\label{eq:V_dot_expanded_comp2}
\end{aligned}
\end{equation}
Using \eqref{eq:V_dot_expanded_comp1} and \eqref{eq:V_dot_expanded_comp2} we can then write \eqref{eq:V_dot_expanded_1} as follows 
\begin{equation}
\begin{aligned}
\dot{V} &= \mathbf{e}_{\text{real}}^\top (A_e^\top P + P A_e) \mathbf{e}_{\text{real}} 
+ \mathbf{e}_{\text{imag}}^\top (A_e^\top P + P A_e) \mathbf{e}_{\text{imag}} \\
&\quad + 2 \lambda \tilde{\theta} \left( \mathbf{e}_{\text{real}}^\top P R_{\text{real}} + \mathbf{e}_{\text{imag}}^\top P R_{\text{imag}} \right)
+ 2 \frac{|\lambda|}{\gamma} \tilde{\theta} \dot{\tilde{\theta}}
\label{eq:V_dot_expanded_2}
\end{aligned}
\end{equation}
using \eqref{eq:Lyap_eq} we can write the first and second terms of \eqref{eq:V_dot_expanded_2}
\begin{equation}
\begin{aligned}
\dot{V} &= - (\mathbf{e}_{\text{real}}^\top Q \mathbf{e}_{\text{real}} 
+ \mathbf{e}_{\text{imag}}^\top Q \mathbf{e}_{\text{imag}}) \\
&\quad + 2 \lambda \tilde{\theta} \left( \mathbf{e}_{\text{real}}^\top P R_{\text{real}} + \mathbf{e}_{\text{imag}}^\top P R_{\text{imag}} \right)
+ 2 \frac{|\lambda|}{\gamma} \tilde{\theta} \dot{\tilde{\theta}}
\label{eq:V_dot_expanded_3}
\end{aligned}
\end{equation}
We can express $(\mathbf{e}_{\text{real}}^\top Q \mathbf{e}_{\text{real}}
+ \mathbf{e}_{\text{imag}}^\top Q \mathbf{e}_{\text{imag}})$ the first term in \eqref{eq:V_dot_expanded_3} as $\bar{\mathbf{e}}^\top Q \mathbf{e} \in \mathbb{R}$  which we will show below
\begin{equation}
    \begin{aligned}
        \label{eq:first_term_Q}
        \bar{\mathbf{e}}^\top Q \mathbf{e} 
          &= (\mathbf{e}_{\text{real}} - i \mathbf{e}_{\text{imag}})^\top Q (\mathbf{e}_{\text{real}} + i \mathbf{e}_{\text{imag}}) \\
          &= (\mathbf{e}_{\text{real}}^\top - i \mathbf{e}_{\text{imag}}^\top)(Q \mathbf{e}_{\text{real}} + i Q \mathbf{e}_{\text{imag}})\\
          &= \mathbf{e}_{\text{real}}^\top Q \mathbf{e}_{\text{real}} - i \mathbf{e}_{\text{imag}}^\top Q \mathbf{e}_{\text{real}} \\ &+ i \mathbf{e}_{\text{real}}^\top Q \mathbf{e}_{\text{imag}} + \mathbf{e}_{\text{imag}}^\top Q \mathbf{e}_{\text{imag}}
    \end{aligned}
\end{equation}
where Q is real and symmetric
\begin{equation}
\begin{aligned}
\label{eq:scalar_transpose_Q}
\mathbf{e}_{\text{real}}^\top Q \mathbf{e}_{\text{imag}} &= (\mathbf{e}_{\text{imag}}^\top Q \mathbf{e}_{\text{real}})^\top = \mathbf{e}_{\text{imag}}^\top Q \mathbf{e}_{\text{real}} \end{aligned}
\end{equation}
which holds similarly as discussed for \eqref{eq:scalar_transpose}. Using \eqref{eq:scalar_transpose_Q} we simplify \eqref{eq:first_term_Q}
\begin{equation}
\begin{aligned}
\bar{\mathbf{e}}^\top Q \mathbf{e} &= \mathbf{e}_{\text{real}}^\top Q \mathbf{e}_{\text{real}} + \mathbf{e}_{\text{imag}}^\top Q \mathbf{e}_{\text{imag}} + i \underbrace{\left( \mathbf{e}_{\text{real}}^\top Q \mathbf{e}_{\text{imag}} - \mathbf{e}_{\text{imag}}^\top Q \mathbf{e}_{\text{real}} \right)}_{0} \\
\bar{\mathbf{e}}^\top Q \mathbf{e} &= \underbrace{\mathbf{e}_{\text{real}}^\top Q \mathbf{e}_{\text{real}} + \mathbf{e}_{\text{imag}}^\top Q \mathbf{e}_{\text{imag}}}_{\bar{\mathbf{e}}^\top Q \mathbf{e}}
\end{aligned}
\label{eq:real_symmetric_Q_simplification}
\end{equation}
so $\bar{\mathbf{e}}^\top Q \mathbf{e} \in \mathbb{R}$ because the imaginary term is equal to 0. 
Now we can look the term $ \left( \mathbf{e}_{\text{real}}^\top P R_{\text{real}} + \mathbf{e}_{\text{imag}}^\top P R_{\text{imag}} \right)$ in \eqref{eq:V_dot_expanded_2} which is equivalent to $\Re(\bar{\mathbf{e}}^\top P R)$ and we will show below
\begin{equation}
\begin{aligned}
\bar{\mathbf{e}}^\top P R 
&= (\mathbf{e}_{\text{real}}^\top - i \mathbf{e}_{\text{imag}}^\top) P (R_{\text{real}} + i R_{\text{imag}}) \\
&= (\mathbf{e}_{\text{real}}^\top - i \mathbf{e}_{\text{imag}}^\top)(P R_{\text{real}} + i P R_{\text{imag}}) \\
&= \mathbf{e}_{\text{real}}^\top P R_{\text{real}} - i \mathbf{e}_{\text{imag}}^\top P R_{\text{real}} 
\\ &+ i \mathbf{e}_{\text{real}}^\top P R_{\text{imag}} - i^2 \mathbf{e}_{\text{imag}}^\top P R_{\text{imag}} \\
&= \underbrace{\mathbf{e}_{\text{real}}^\top P R_{\text{real}} + \mathbf{e}_{\text{imag}}^\top P R_{\text{imag}}}_{\Re\left( \bar{\mathbf{e}}^\top P R \right)}
\\& + i \underbrace{\left( \mathbf{e}_{\text{real}}^\top P R_{\text{imag}} - \mathbf{e}_{\text{imag}}^\top P R_{\text{real}} \right)}_{\Im\left( \bar{\mathbf{e}}^\top P R \right)}. \\
\end{aligned}
\label{eq:e_conj_PR_expansion}
\end{equation}
Using \eqref{eq:e_conj_PR_expansion} we can no rewrite \eqref{eq:V_dot_expanded_3} as
\begin{equation}
\begin{aligned}
\dot{V} &= -\bar{\mathbf{e}}^\top Q \mathbf{e} + 2 \lambda \tilde{\theta} \Re(\bar{\mathbf{e}}^\top P R) + 2 \frac{|\lambda|}{\gamma} \tilde{\theta} \dot{\tilde{\theta}}.
\end{aligned}
\label{eq:V_dot_final}
\end{equation}
We choose the adaptive laws so that the the second term in \eqref{eq:V_dot_final} vanishes
\begin{equation}
\begin{aligned}
\dot{V} &= -\bar{\mathbf{e}}^\top Q \mathbf{e} + 2 |\lambda| \tilde{\theta} (\text{sign}(\lambda)) \Re(\bar{\mathbf{e}}^\top P R) + \frac{2}{\gamma} |\lambda| \tilde{\theta} \dot{\tilde{\theta}} \\
\dot{\tilde{\theta}} &= -\text{sign}(\lambda)   \gamma   \Re(\bar{\mathbf{e}}^\top P R) \\
\dot{V} &= -\bar{\mathbf{e}}^\top Q \mathbf{e} + 2 |\lambda| \tilde{\theta} (  \text{sign}(\lambda) \Re(\bar{\mathbf{e}}^\top P R)
\\ &- \frac{\gamma}{\gamma}   \text{sign}(\lambda)    \Re(\bar{\mathbf{e}}^\top P R)) \\
\dot{V} &= -\bar{\mathbf{e}}^\top Q \mathbf{e}
\end{aligned}
\label{eq:Lyapunov_analysis}
\end{equation}
where $\bar{\mathbf{e}}^\top Q \mathbf{e} \in \mathbb{R}$ as shown in equations \eqref{eq:first_term_Q} - \eqref{eq:real_symmetric_Q_simplification}.

\subsection{Adaptive Control With Input Constraints}
\label{sec:app_ac_control}
Using \eqref{eq:updated_plant}, \eqref{eq:degraded_reference_model}, \eqref{eq:lambdahat_control}, \eqref{eq:heading_sat}, and \eqref{eq:u2sat_rewritten} we obtain the following for the velocity error dynamics, 
\begin{equation}\label{eq:velocity_error_revisited}
\begin{aligned}
\dot{e}_v = \dot{v}_a - \dot{v}_{\text{ref}}\\
= i \lambda u_2^{sat} V_a e^{i\psi} -  i \dot{\psi}_{\text{ref}} V_{\text{ref}} e^{i\psi_{\text{ref}}}- i V_a \hat{\lambda} \Delta_{\mathrm{sat}} e^{i\psi}
\\= i \lambda\bigl(u_2 + \Delta_{sat}\bigr) V_a e^{i\psi}
- i u_{2\text{ref}} V_{\text{ref}} e^{i\psi_{\text{ref}}}
- i V_a \hat{\lambda} \Delta_{\mathrm{sat}} e^{i\psi}
\\= i \lambda u_2 V_a e^{i\psi} + i \lambda \Delta_{sat} V_a e^{i\psi} - i u_{2\text{ref}} V_{a} e^{i\psi_{\text{ref}}} - i V_a \hat{\lambda} \Delta_{\mathrm{sat}} e^{i\psi}
\\=
i V_a\Bigl(\lambda u_2 e^{i\psi}-u_{2\text{ref}} e^{i\psi_{\text{ref}}}\Bigr)
 + 
(\lambda - \hat{\lambda}) i V_a \Delta_{sat} e^{i\psi}.
\end{aligned}
\end{equation}

Substituting \eqref{eq:lambdahat_control} into \eqref{eq:velocity_error_revisited} we obtain \begin{equation}\label{eq:uncertainty_velocity_error}
\begin{aligned}
\dot{e}_v = i V_a
\Bigl(\lambda 
\frac{1}{\hat{\lambda} e^{i\psi}}
\bigl[
\tfrac{\mathbf{k}^\top \mathbf{e}}{i V_a}+u_{2\text{ref}} e^{i\psi_{\text{ref}}}
\bigr]
e^{i\psi}
- 
u_{2\text{ref}} e^{i\psi_{\text{ref}}}\Bigr) \\ + (\lambda - \hat{\lambda}) i V_a \Delta_{sat} e^{i\psi}\\ =
i V_a\Bigl(\lambda 
\frac{1}{\hat{\lambda}}
\bigl[
\tfrac{\mathbf{k}^\top \mathbf{e}}{i V_a}+u_{2\text{ref}} e^{i\psi_{\text{ref}}}
\bigr]
- 
u_{2\text{ref}} e^{i\psi_{\text{ref}}}\Bigr)
\\+ (\lambda - \hat{\lambda}) i V_a \Delta_{sat} e^{i\psi}\\
=
i V_a\biggl[
\frac{\lambda}{\hat{\lambda}}
\tfrac{\mathbf{k}^\top \mathbf{e}}{i V_a}
+ 
\frac{\lambda}{\hat{\lambda}} u_{2\text{ref}} e^{i\psi_{\text{ref}}}
- 
u_{2\text{ref}} e^{i\psi_{\text{ref}}}
\biggr]
\\+
(\lambda - \hat{\lambda})
 i V_a \Delta_{sat} e^{i\psi}\\
=
\frac{\lambda}{\hat{\lambda}} \mathbf{k}^\top \mathbf{e}
+ 
\Bigl(\frac{\lambda}{\hat{\lambda}} - 1\Bigr)
i V_a u_{2\text{ref}} e^{i\psi_{\text{ref}}}
\\+ 
(\lambda - \hat{\lambda})
 i V_a \Delta_{sat} e^{i\psi}
 \neq \delta.
\end{aligned}
\end{equation}

Next we define 
\begin{equation}
\hat{\theta}  = \frac{1}{\hat{\lambda}},
\quad
\theta^*  = \frac{1}{\lambda^*},
\quad
\tilde{\theta}  = \hat{\theta}  - \theta^*.
\label{eq:theta_defintion}
\end{equation}
\noindent therefore
\begin{equation}
\lambda^* \hat{\theta} - 1
 = 
\lambda^*\tilde{\theta}.
\label{eq:lambda_tilde_simplification}
\end{equation}
Using \eqref{eq:theta_defintion} and \eqref{eq:lambda_tilde_simplification} in \eqref{eq:uncertainty_velocity_error}:
\begin{equation}\label{eq:theta_instead}
\begin{aligned}
\dot{e}_v &= 
\frac{\lambda}{\hat{\lambda}} \mathbf{k}^\top \mathbf{e}
+\Bigl(\frac{\lambda}{\hat{\lambda}} - 1\Bigr) i V_a u_{2\text{ref}} e^{i\psi_{\text{ref}}}
+ 
(\lambda-\hat{\lambda}) i V_a \Delta_{sat} e^{i\psi}\\
&=
\mathbf{k}^\top \mathbf{e}
+ 
\bigl(\lambda\tilde{\theta}\bigr) \mathbf{k}^\top \mathbf{e}
+ 
\bigl(\lambda\tilde{\theta}\bigr) i V_a u_{2\text{ref}} e^{i\psi_{\text{ref}}}
- 
\tilde{\lambda} i V_a \Delta_{sat} e^{i\psi}\\
&=
\mathbf{k}^\top \mathbf{e}
+ 
\lambda\tilde{\theta}\Bigl(\mathbf{k}^\top \mathbf{e} + i V_a u_{2\text{ref}} e^{i\psi_{\text{ref}}}\Bigr)
- 
\tilde{\lambda} i V_a \Delta_{sat} e^{i\psi}.
\end{aligned}
\end{equation}
Thus, if $\hat{\theta}\neq \theta$, extra terms remain, and $\tilde{\lambda}$ also adds a residual term when $\Delta_{sat}\neq0$.

\subsubsection{Lyapunov-Based Analysis and Adaptive Law Derivation}

Without uncertainty or saturation, the error dynamics were:
\begin{equation}\label{eq:error_dynamics_1}
\dot{\mathbf{e}} = 
\begin{bmatrix}
\dot{e}_I\\
\dot{e}_r\\
\dot{e}_v
\end{bmatrix}
=
\underbrace{
\begin{bmatrix}
0 & 1 & 0\\
0 & 0 & 1\\
-k_I & -k_P & -k_D
\end{bmatrix}
}_{A_e}
\begin{bmatrix}
e_I\\
e_r\\
e_v
\end{bmatrix}
=
A_e \mathbf{e}.
\end{equation}
With uncertainty, extra terms appear which we denote $R$
\begin{equation}\label{eq:error_dynamics_uncertainty}
\begin{aligned}
\dot{\mathbf{e}} =
\underbrace{\begin{bmatrix}0 & 1 & 0\\0 & 0 & 1\\-k_I & -k_P & -k_D\end{bmatrix}}_{A_e}
\begin{bmatrix}e_I\\ e_r\\ e_v\end{bmatrix}
\\+
\lambda \tilde{\theta}\Bigl(
\underbrace{
\begin{bmatrix}
0 & 0 & 0\\
0 & 0 & 0\\
-k_I & -k_P & -k_D
\end{bmatrix}
\begin{bmatrix}
e_I\\ e_r\\ e_v
\end{bmatrix} +
i V_a u_{2\text{ref}} e^{i\psi_{\text{ref}}}
\begin{bmatrix}0\\0\\1\end{bmatrix}
}_{R}
\Bigr).
\end{aligned}
\end{equation}
which can be written in compact form
\begin{equation}\label{eq:compact_error_uncertainty}
\dot{\mathbf{e}} = A_e \mathbf{e} + \lambda \tilde{\theta} R.
\end{equation}
With saturation, we get an additional term denoted by $S$
\begin{equation}\label{eq:error_dynamics_uncertainty_saturation}
\begin{aligned}
\dot{\mathbf{e}} =
\begin{bmatrix}0 & 1 & 0\\0 & 0 & 1\\-k_I & -k_P & -k_D\end{bmatrix}
\begin{bmatrix}e_I\\ e_r\\ e_v\end{bmatrix}
\\+
\lambda \tilde{\theta}
\Bigl(
\begin{bmatrix}0 & 0 & 0\\0 & 0 & 0\\-k_I & -k_P & -k_D\end{bmatrix}
\begin{bmatrix}e_I\\ e_r\\ e_v\end{bmatrix}
+
i V_a u_{2\text{ref}} e^{i\psi_{\text{ref}}}
\begin{bmatrix}0\\0\\1\end{bmatrix}
\Bigr)
\\
\quad\quad - \tilde{\lambda}
\underbrace{
i V_a \Delta_{sat} e^{i\psi}
\begin{bmatrix}0\\0\\1\end{bmatrix}
}_{S}.
\end{aligned}
\end{equation}
which can be written compactly as
\begin{equation}\label{eq:compact_error_uncertainty_saturation}
\dot{\mathbf{e}} = A_e \mathbf{e} + \lambda \tilde{\theta} R - \tilde{\lambda} S.
\end{equation}

\subsubsection{Adaptive Law With Input Constraints}
We consider error dynamics that include both parametric uncertainty and control saturation. The error dynamics are:
\begin{equation}\label{eq:error_dynamics_sat}
\dot{\mathbf{e}} = A_e \mathbf{e} + \lambda \tilde{\theta} R - \tilde{\lambda} S
\end{equation}
where
\begin{itemize}
\item $A_e \in \mathbb{R}^{3\times3}$ is a Hurwitz (stable) real matrix (so that $A_e^H = A_e^\top$). In fact $A_e$ is chosen of the form,
\begin{equation}
A_e = \begin{bmatrix} 0 & 1 & 0 \\ 0 & 0 & 1 \\ -k_I & -k_P & -k_D \end{bmatrix}.
\end{equation}
with appropriate gains $k_P, k_I, k_D$ $\in \mathbb{R}$. 
\item $\lambda \in \mathbb{R}$, $\tilde{\theta} \in \mathbb{R}$, $R \in \mathbb{C}^{3}$, $\tilde{\lambda} \in \mathbb{R}$), $S \in \mathbb{C}^{3}$.
\end{itemize}

We choose the following Lyapunov candidate:
\begin{equation}\label{eq:V_candidate}
V(\mathbf{e},\tilde{\theta},\tilde{\lambda}) = \bar{\mathbf{e}}^\top P \mathbf{e} + \frac{|\lambda|}{\gamma_{\theta}} \tilde{\theta}^2 + \frac{1}{\gamma_{\lambda}} \tilde{\lambda}^2
\end{equation}
where
\begin{itemize}
\item $P \in \mathbb{R}^{3\times3}$ is a symmetric positive-definite matrix chosen to satisfy \eqref{eq:Lyap_eq}
\item $\gamma_{\theta}>0$ and $\gamma_{\lambda}>0$ are adaptation gains.
\end{itemize}

Next we differentiate $V$ similarly to \eqref{eq:V_dot_derivative} and we obtain the following
\begin{equation}\label{eq:V_dot_subs}
\begin{aligned}
    \dot{V} 
    =
    2 \mathbf{e}_{\text{real}}^\top P \dot{\mathbf{e}}_{\text{real}} 
    + 2 \mathbf{e}_{\text{imag}}^\top P \dot{\mathbf{e}}_{\text{imag}} 
    + 2 \frac{|\lambda|}{\gamma} \tilde{\theta} \dot{\tilde{\theta}} + \frac{2}{\gamma}\tilde{\lambda}\dot{\tilde{\lambda}}
    \end{aligned}
\end{equation}
We simplify \eqref{eq:V_dot_subs} following similar steps as in \eqref{eq:V_dot_expanded_1} - \eqref{eq:e_conj_PR_expansion} and we obtain
\begin{equation}\label{eq:V_dot_final_before_adapt}
\begin{aligned}    
\dot{V} = -\bar{\mathbf{e}}^\top Q \mathbf{e} + 2\lambda \tilde{\theta} \Re\{\bar{\mathbf{e}}^\top P R\} - 2\tilde{\lambda} \Re\{\bar{\mathbf{e}}^\top P S\} \\+ \frac{2|\lambda|}{\gamma_{\theta}} \tilde{\theta} \dot{\tilde{\theta}} + \frac{2}{\gamma_{\lambda}} \tilde{\lambda} \dot{\tilde{\lambda}}.
\end{aligned}
\end{equation}
We then group terms as follows
\begin{equation}\label{eq:V_dot_grouped}
\begin{aligned}
\dot{V} = -\bar{\mathbf{e}}^\top Q \mathbf{e} \\+ 2|\lambda| \tilde{\theta}\Bigl(\text{sign}(\lambda) \Re\{\bar{\mathbf{e}}^\top P R\} + \frac{1}{\gamma_{\theta}}\dot{\tilde{\theta}}\Bigr) \\+ 2\tilde{\lambda}\Bigl(-\Re\{\bar{\mathbf{e}}^\top P S\} + \frac{1}{\gamma_{\lambda}}\dot{\tilde{\lambda}}\Bigr).
\end{aligned}
\end{equation}

We choose the adaptive laws so that the the second and third terms in \eqref{eq:V_dot_grouped} vanish
\begin{equation}\label{eq:adaptive_laws_app}
\dot{\tilde{\theta}} = -\gamma_{\theta} \text{sign}(\lambda) \Re\{\bar{\mathbf{e}}^\top P R\},\quad
\dot{\tilde{\lambda}} = \gamma_{\lambda} \Re\{\bar{\mathbf{e}}^\top P S\}.
\end{equation}
Substituting into \eqref{eq:V_dot_grouped} we obtain
\begin{equation}\label{eq:V_dot_final_app}
\dot{V} = -\bar{\mathbf{e}}^\top Q \mathbf{e}.
\end{equation}
Since $Q$ is positive definite, $\dot{V} \le 0$.  By Barbalat's Lemma, $\dot{V}(t)\to 0$ as $t\to\infty$, which implies
\begin{equation}
-\mathbf{e}(t)^\top Q \mathbf{e}(t) \to 0 \quad \Longrightarrow \quad \mathbf{e}(t) \to 0.
\end{equation}

\subsection{Ways in which $R_{\text{ref}}$ can be selected}
\label{app:choose_R_ref}
\textbf{1)} Selecting $R_{\text{ref}} = R_{\min}$
\begin{equation}
|\dot{\psi}_{\text{ref}}| = \frac{V_{\text{ref}}}{R_{\text{ref}}} = \frac{V_a}{R_{\min}} = |\dot{\psi}_\text{max}|
\end{equation}
This means the path is chosen such that each turn would require making the tightest turn that the DV is capable of.

\textbf{2}) By selecting $R_{\text{ref}} > R_{\min}$
\begin{equation}
|\dot{\psi}_{\text{ref}}| = \frac{V_{\text{ref}}}{R_{\text{ref}}} < \frac{V_a}{R_{\min}} = |\dot{\psi}_\text{max}|
\end{equation}
The path is chosen such that each turn requires a turning rate less than the maximum possible by the DV.

\textbf{3}) By selecting $R_{\text{ref}} < R_{\min}$
\begin{equation}
|\dot{\psi}_{\text{ref}}| = \frac{V_{\text{ref}}}{R_{\text{ref}}} > \frac{V_a}{R_{\min}} = |\dot{\psi}_\text{max}|
\end{equation}
The path is chosen such that each turn requires a turning rate greater than what the DV is capable of. \textit{Note: } If we pick \#3, no matter how good of a controller, the DV cannot follow the path because the path demands a required rate greater than what the DV is capable of. By picking \#1 or \#2 we ensures that $ \dot{\psi}_{\mathrm{ref}} \in [\dot{\psi}_{\min}, \dot{\psi}_{\max}] $. 

But this is only possible if we have perfect knowledge of the DV capabilities. In a turning rate LOE scenario where $ $ and $ \epsilon > 0 $, the turning rate capability becomes compromised in the following way:
\begin{equation}
|\dot{\psi}_{\max}| \Longrightarrow \lambda|\dot{\psi}_{\max}|, \quad \lambda \in [\epsilon, 1],  \quad \epsilon > 0
\end{equation}
\begin{equation} \begin{aligned}
 \lambda |\dot{\psi}_{\max}| =  \lambda \frac{V_a}{R_{\min}} =  \frac{V_a}{\frac{1}{\lambda}R_{\min}}
\end{aligned}
\end{equation}
So if we picked (1) $R_{\mathrm{ref}} = R_{\min}$
\begin{equation}
\dot{\psi}_{\mathrm{ref}} = \frac{V_{\mathrm{ref}}}{R_{\mathrm{ref}}} = \frac{V_{\mathrm{ref}}}{R_{\min}}
\end{equation}
\textbf{Using (1)} Path is chosen such that each turn would require making the sharpest turn that is possible by the DV:
\begin{equation}
\dot{\psi}_{\mathrm{ref}} = \left|\dot{\psi}_{\max}\right|,
\end{equation}
but the DV turning rate capability is:
\begin{equation}
\dot{\psi}_{\mathrm{ref}} = \lambda \left|\psi_{\max}\right|
\end{equation}

\begin{equation}
\dot{\psi}_{\mathrm{ref}} = \frac{V_{\mathrm{ref}}}{R_{\mathrm{ref}}} > \frac{V_{\mathrm{ref}}}{\frac{1}{\lambda} R_{\min}}
\end{equation}
This means that each turn requires a turning rate greater than what the compromised DV can do:
\begin{equation}
\dot{\psi}_{\mathrm{ref}} > \lambda \left|\psi_{\max}\right|
\end{equation}

\textbf{Using (2)} The path is chosen such that $R_{\mathrm{ref}} > R_{\min}$
\begin{equation}
\dot{\psi}_{\mathrm{ref}} = \frac{V_{\mathrm{ref}}}{R_{\mathrm{ref}}} < \frac{V_{\mathrm{ref}}}{R_{\min}}
\end{equation}
The results depend on $R_{\mathrm{ref}}$,
\newline \textbf{a)} If 
\begin{equation}
    R_{\text{min}} < R_{\text{ref}} < \frac{1}{\lambda} R_{\text{min}}
\end{equation}
each turn requires a turning rate greater than what the compromised DV is capable of.
\newline\textbf{b)} If 
\begin{equation}
    R_{\text{min}} < \frac{1}{\lambda} R_{\text{min}} < R_{\text{ref}}  
\end{equation}
each turn requires a turning rate less than the maximum turning rate of the compromised DV.
\newline\textbf{c)} If 
\begin{equation}
    R_{\text{min}} <  \frac{1}{\lambda} R_{\text{min}} = R_{\text{ref}} 
\end{equation}
each turn requires a turning equal to max turning rate of the compromised DV.
Therefore by picking $R_{\text{ref}}$ such that 
\begin{equation}
    R_{\text{ref}} = \frac{1}{\lambda_{\text{min}}}R_{\text{min}}
\end{equation}
where $\lambda_{\text{min}} < \lambda$ is the worst case LOE, ensures the path generated by the reference model does require a turning rate that exceeds the capabilities of the DV. 
\end{document}